 \documentclass[twocolumn, trackchanges]{aastex63}
 
 \usepackage{newtxtext,newtxmath}
 \usepackage{graphicx}	% Including figure files
 \usepackage{amsmath}	% Advanced maths commands
 \usepackage{amssymb}	% Extra maths symbols
 \usepackage{rotating}
 \usepackage{multirow}

\usepackage{xcolor}

\usepackage{lineno}
\usepackage[utf8]{inputenc}
\DeclareUnicodeCharacter{2212}{-}
\received{2021 March 8}
\revised{2021 April 7}
\accepted{2021 April 7}
\published{2021 August 9}
\submitjournal{ApJL}

\shorttitle{TeV emission of Galactic plane sources with HAWC and H.E.S.S.}
\shortauthors{H.E.S.S. and HAWC collaborations}
\graphicspath{{./}{figures/}}
\begin{document}

\title{\Large TeV emission of Galactic plane sources with HAWC and H.E.S.S.}

\correspondingauthor{Armelle Jardin-Blicq}
\email{armelle.jardin-blicq@mpi-hd.mpg.de}

\correspondingauthor{Vincent Marandon, François Brun}
\email{contact.hess@hess-experiment.eu}

\author{H.~Abdalla} 
\affiliation{University of Namibia, Department of Physics, Private Bag 13301, Windhoek 10005, Namibia}

\author[0000-0003-1157-3915]{F.~Aharonian} 
\affiliation{Dublin Institute for Advanced Studies, 31 Fitzwilliam Place, Dublin 2, Ireland}
\affiliation{Max-Planck Institute for Nuclear Physics, D-69117 Heidelberg, Germany}
\affiliation{High Energy Astrophysics Laboratory, RAU,  123 Hovsep Emin Street  Yerevan 0051, Armenia}

\author{F.~Ait~Benkhali} 
\affiliation{Max-Planck Institute for Nuclear Physics, D-69117 Heidelberg, Germany}

\author{E.O.~Ang\"uner} 
\affiliation{Aix Marseille Universit\'e, CNRS/IN2P3, CPPM, Marseille, France}

\author{C.~Arcaro} 
\affiliation{Centre for Space Research, North-West University, Potchefstroom 2520, South Africa}

\author{C.~Armand} 
\affiliation{Laboratoire d'Annecy de Physique des Particules, Université Grenoble Alpes, Université Savoie Mont Blanc, CNRS, LAPP, F-74000 Annecy, France}

\author[0000-0001-5067-2620]{T.~Armstrong}
\affiliation{University of Oxford, Department of Physics, Denys Wilkinson Building, Keble Road, Oxford OX1 3RH, UK}

\author[0000-0002-2153-1818]{H.~Ashkar} 
\affiliation{IRFU, CEA, Universit\'e Paris-Saclay, F-91191 Gif-sur-Yvette, France}

\author[0000-0002-9326-6400]{M.~Backes}
\affiliation{University of Namibia, Department of Physics, Private Bag 13301, Windhoek 10005, Namibia}
\affiliation{Centre for Space Research, North-West University, Potchefstroom 2520, South Africa}

\author{V.~Baghmanyan} 
\affiliation{Instytut Fizyki J\c{a}drowej PAN, ul. Radzikowskiego 152, 31-342 Krak{\'o}w, Poland}

% ...

\collaboration{233}{(H.E.S.S. collaboration)}

\author[0000-0003-0197-5646]{A.~Albert}
\affiliation{Physics Division, Los Alamos National Laboratory, Los Alamos, NM, USA }

\author[0000-0001-8749-1647]{R.~Alfaro}
\affiliation{Instituto de F\'{i}sica, Universidad Nacional Autónoma de México, Ciudad de Mexico, Mexico }

\author{C.~Alvarez}
\affiliation{Universidad Autónoma de Chiapas, Tuxtla Gutiérrez, Chiapas, Mexico}

\author{J.C.~Arteaga-Velázquez}
\affiliation{Universidad Michoacana de San Nicolás de Hidalgo, Morelia, Mexico }

\author[0000-0002-3032-663X]{K.P.~Arunbabu}
\affiliation{Instituto de Geof\'{i}sica, Universidad Nacional Autónoma de México, Ciudad de Mexico, Mexico }

\author[0000-0002-4020-4142]{D.~Avila Rojas}
\affiliation{Instituto de F\'{i}sica, Universidad Nacional Autónoma de México, Ciudad de Mexico, Mexico }

\author[0000-0003-0477-1614]{V.~Baghmanyan}
\affiliation{Institute of Nuclear Physics Polish Academy of Sciences, PL-31342 IFJ-PAN, Krakow, Poland }

\author[0000-0003-3207-105X]{E.~Belmont-Moreno}
\affiliation{Instituto de F\'{i}sica, Universidad Nacional Autónoma de México, Ciudad de Mexico, Mexico }

\author[0000-0001-5537-4710]{S.Y.~BenZvi}
\affiliation{Department of Physics \& Astronomy, University of Rochester, Rochester, NY , USA }

\author[0000-0002-5493-6344]{C.~Brisbois}
\affiliation{Department of Physics, University of Maryland, College Park, MD, USA }

\author[0000-0002-4042-3855]{K.S.~Caballero-Mora}
\affiliation{Universidad Autónoma de Chiapas, Tuxtla Gutiérrez, Chiapas, Mexico}

\author[0000-0003-2158-2292]{T.~Capistrán}
\affiliation{Instituto de Astronom\'{i}a, Universidad Nacional Autónoma de México, Ciudad de Mexico, Mexico }

\author[0000-0002-8553-3302]{A.~Carramiñana}
\affiliation{Instituto Nacional de Astrof\'{i}sica, Óptica y Electrónica, Puebla, Mexico }

\author[0000-0002-6144-9122]{S.~Casanova}
\altaffiliation{Author in both collaborations}
\affiliation{Institute of Nuclear Physics Polish Academy of Sciences, PL-31342 IFJ-PAN, Krakow, Poland }

\author[0000-0002-7607-9582]{U.~Cotti}
\affiliation{Universidad Michoacana de San Nicolás de Hidalgo, Morelia, Mexico }

\author[0000-0002-1132-871X]{J.~Cotzomi}
\affiliation{Facultad de Ciencias F\'{i}sico Matemáticas, Benemérita Universidad Autónoma de Puebla, Puebla, Mexico}

\author[0000-0002-7747-754X]{S.~Coutiño de León}
\affiliation{Instituto Nacional de Astrof\'{i}sica, Óptica y Electrónica, Puebla, Mexico }

\author[0000-0001-9643-4134]{E.~De la Fuente}
\affiliation{Departamento de F\'{i}sica, Centro Universitario de Ciencias Exactase Ingenierias, Universidad de Guadalajara, Guadalajara, Mexico }
\affiliation{Institute for Cosmic Ray Research, University of Tokyo, 277-8582 Chiba, Kashiwa, Kashiwanoha, 5 Chome-1-5, Japan}

\author[0000-0002-8528-9573]{C.~de León}
\affiliation{Universidad Michoacana de San Nicolás de Hidalgo, Morelia, Mexico }

\author[0000-0001-8487-0836]{R.~Diaz Hernandez}
\affiliation{Instituto Nacional de Astrof\'{i}sica, Óptica y Electrónica, Puebla, Mexico }

\author[0000-0002-0087-0693]{J.C.~Díaz-Vélez}
\affiliation{Departamento de F\'{i}sica, Centro Universitario de Ciencias Exactase Ingenierias, Universidad de Guadalajara, Guadalajara, Mexico }

\author[0000-0001-8451-7450]{B.L.~Dingus}
\affiliation{Physics Division, Los Alamos National Laboratory, Los Alamos, NM, USA }

\author[0000-0002-2987-9691]{M.A.~DuVernois}
\affiliation{Department of Physics, University of Wisconsin-Madison, Madison, WI, USA }

\author[0000-0003-2169-0306]{M.~Durocher}
\affiliation{Physics Division, Los Alamos National Laboratory, Los Alamos, NM, USA }

\author[0000-0003-2338-0344]{R.W.~Ellsworth}
\affiliation{Department of Physics, University of Maryland, College Park, MD, USA }

\author[0000-0001-5737-1820]{K.~Engel}
\affiliation{Department of Physics, University of Maryland, College Park, MD, USA }

\author[0000-0001-7074-1726]{C.~Espinoza}
\affiliation{Instituto de F\'{i}sica, Universidad Nacional Autónoma de México, Ciudad de Mexico, Mexico }

\author{K.L.~Fan}
\affiliation{Department of Physics, University of Maryland, College Park, MD, USA }

\author{M.~Fernández Alonso}
\affiliation{Department of Physics, Pennsylvania State University, University Park, PA, USA }

\author{N.~Fraija}
\affiliation{Instituto de Astronom\'{i}a, Universidad Nacional Autónoma de México, Ciudad de Mexico, Mexico }

\author{A.~Galván-Gámez}
\affiliation{Instituto de Astronom\'{i}a, Universidad Nacional Autónoma de México, Ciudad de Mexico, Mexico }

\author{D.~Garcia}
\affiliation{Instituto de F\'{i}sica, Universidad Nacional Autónoma de México, Ciudad de Mexico, Mexico }

\author[0000-0002-4188-5584]{J.A.~García-González}
\affiliation{Tecnologico de Monterrey, Escuela de Ingeniería y Ciencias, Ave. Eugenio Garza Sada 2501, Monterrey, N.L., 64849, Mexico }

\author[0000-0003-1122-4168]{F.~Garfias}
\affiliation{Instituto de Astronom\'{i}a, Universidad Nacional Autónoma de México, Ciudad de Mexico, Mexico }

\author[0000-0001-9745-5738]{G.~Giacinti}
\altaffiliation{Author in both collaborations}
\affiliation{Max-Planck Institute for Nuclear Physics, D-69117 Heidelberg, Germany}

\author[0000-0002-5209-5641]{M.M.~González}
\affiliation{Instituto de Astronom\'{i}a, Universidad Nacional Autónoma de México, Ciudad de Mexico, Mexico }

\author[0000-0002-9790-1299]{J.A.~Goodman}
\affiliation{Department of Physics, University of Maryland, College Park, MD, USA }

\author[0000-0001-9844-2648]{J.P.~Harding}
\affiliation{Physics Division, Los Alamos National Laboratory, Los Alamos, NM, USA }

\author[0000-0002-2565-8365]{S.~Hernandez}
\affiliation{Instituto de F\'{i}sica, Universidad Nacional Autónoma de México, Ciudad de Mexico, Mexico }

\author{B.~Hona}
\affiliation{Department of Physics and Astronomy, University of Utah, Salt Lake City, UT, USA }

\author[0000-0002-5447-1786]{D.~Huang}
\affiliation{Department of Physics, Michigan Technological University, Houghton, MI, USA }

\author[0000-0002-5527-7141]{F.~Hueyotl-Zahuantitla}
\affiliation{Universidad Autónoma de Chiapas, Tuxtla Gutiérrez, Chiapas, Mexico}

\author{P.~Hüntemeyer}
\affiliation{Department of Physics, Michigan Technological University, Houghton, MI, USA }

\author[0000-0001-5811-5167]{A.~Iriarte}
\affiliation{Instituto de Astronom\'{i}a, Universidad Nacional Autónoma de México, Ciudad de Mexico, Mexico }

\author[0000-0002-6738-9351]{A.~Jardin-Blicq}
\altaffiliation{Author in both collaborations}
\affiliation{Max-Planck Institute for Nuclear Physics, D-69117 Heidelberg, Germany}
\affiliation{Department of Physics, Faculty of Science, Chulalongkorn University, 254 Phayathai Road, Pathumwan, Bangkok 10330, Thailand}
\affiliation{National Astronomical Research Institute of Thailand (Public Organization), Don Kaeo, MaeRim, Chiang Mai 50180, Thailand}

\author[0000-0003-4467-3621]{V.~Joshi}
\altaffiliation{Author in both collaborations}
\affiliation{Friedrich-Alexander-Universit\"at Erlangen-N\"urnberg, Erlangen Centre for Astroparticle Physics, Erwin-Rommel-Stra\ss e 1, D-91058 Erlangen, Germany}

\author[0000-0003-4785-0101]{D.~Kieda}
\affiliation{Department of Physics and Astronomy, University of Utah, Salt Lake City, UT, USA }

\author[0000-0002-2467-5673]{W.H.~Lee}
\affiliation{Instituto de Astronom\'{i}a, Universidad Nacional Autónoma de México, Ciudad de Mexico, Mexico }

\author[0000-0001-5516-4975]{H.~León Vargas}
\affiliation{Instituto de F\'{i}sica, Universidad Nacional Autónoma de México, Ciudad de Mexico, Mexico }

\author[0000-0003-2696-947X]{J.T.~Linnemann}
\affiliation{Department of Physics and Astronomy, Michigan State University, East Lansing, MI, USA }

\author[0000-0001-8825-3624]{A.L.~Longinotti}
\affiliation{Instituto Nacional de Astrof\'{i}sica, Óptica y Electrónica, Puebla, Mexico }

\author[0000-0003-2810-4867]{G.~Luis-Raya}
\affiliation{Universidad Politecnica de Pachuca, Pachuca, Hgo, Mexico }

\author[0000-0002-3882-9477]{R.~López-Coto}
\affiliation{INFN and Universita di Padova, via Marzolo 8, I-35131,Padova,Italy}

\author[0000-0001-8088-400X]{K.~Malone}
\affiliation{Physics Division, Los Alamos National Laboratory, Los Alamos, NM, USA }

\author[0000-0001-9052-856X]{O.~Martinez}
\affiliation{Facultad de Ciencias F\'{i}sico Matemáticas, Benemérita Universidad Autónoma de Puebla, Puebla, Mexico }

\author[0000-0001-9035-1290]{I.~Martinez-Castellanos}
\affiliation{Department of Physics, University of Maryland, College Park, MD, USA }

\author[0000-0002-2824-3544]{J.~Martínez-Castro}
\affiliation{Centro de Investigaci\'on en Computaci\'on, Instituto Polit\'ecnico Nacional, M\'exico City, Mexico.}

\author[0000-0002-2610-863X]{J.A.~Matthews}
\affiliation{Dept of Physics and Astronomy, University of New Mexico, Albuquerque, NM, USA }

\author[0000-0002-8390-9011]{P.~Miranda-Romagnoli}
\affiliation{Universidad Autónoma del Estado de Hidalgo, Pachuca, Mexico }

\author{J.A.~Morales-Soto}
\affiliation{Universidad Michoacana de San Nicolás de Hidalgo, Morelia, Mexico }

\author[0000-0002-1114-2640]{E.~Moreno}
\affiliation{Facultad de Ciencias F\'{i}sico Matemáticas, Benemérita Universidad Autónoma de Puebla, Puebla, Mexico }

\author[0000-0002-7675-4656]{M.~Mostafá}
\affiliation{Department of Physics, Pennsylvania State University, University Park, PA, USA }

\author[0000-0003-0587-4324]{A.~Nayerhoda}
\affiliation{Institute of Nuclear Physics Polish Academy of Sciences, PL-31342 IFJ-PAN, Krakow, Poland }

\author[0000-0003-1059-8731]{L.~Nellen}
\affiliation{Instituto de Ciencias Nucleares, Universidad Nacional Autónoma de Mexico, Ciudad de Mexico, Mexico }

\author[0000-0001-9428-7572]{M.~Newbold}
\affiliation{Department of Physics and Astronomy, University of Utah, Salt Lake City, UT, USA }

\author[0000-0002-6859-3944]{M.U.~Nisa}
\affiliation{Department of Physics and Astronomy, Michigan State University, East Lansing, MI, USA }

\author[0000-0001-7099-108X]{R.~Noriega-Papaqui}
\affiliation{Universidad Autónoma del Estado de Hidalgo, Pachuca, Mexico }

\author[0000-0002-5448-7577]{N.~Omodei}
\affiliation{Department of Physics, Stanford University: Stanford, CA 94305–4060, USA}

\author{A.~Peisker}
\affiliation{Department of Physics and Astronomy, Michigan State University, East Lansing, MI, USA }

\author[0000-0002-8774-8147]{Y.~Pérez Araujo}
\affiliation{Instituto de Astronom\'{i}a, Universidad Nacional Autónoma de México, Ciudad de Mexico, Mexico }

\author[0000-0001-5998-4938]{E.G.~Pérez-Pérez}
\affiliation{Universidad Politecnica de Pachuca, Pachuca, Hgo, Mexico }

\author[0000-0002-6524-9769]{C.D.~Rho}
\affiliation{Natural Science Research Institute, University of Seoul, Seoul, Republic Of Korea}

\author[0000-0003-1327-0838]{D.~Rosa-González}
\affiliation{Instituto Nacional de Astrof\'{i}sica, Óptica y Electrónica, Puebla, Mexico }

\author[0000-0001-6939-7825]{E.~Ruiz-Velasco}
\altaffiliation{Author in both collaborations}
\affiliation{Max-Planck Institute for Nuclear Physics, D-69117 Heidelberg, Germany}

\author[0000-0002-8610-8703]{F.~Salesa Greus}
\affiliation{Institute of Nuclear Physics Polish Academy of Sciences, PL-31342 IFJ-PAN, Krakow, Poland }
\affiliation{Instituto de F\'{i}sica Corpuscular, CSIC, Universitat de Val\`{e}ncia, E-46980, Paterna, Valencia, Spain}

\author[0000-0001-6079-2722]{A.~Sandoval}
\affiliation{Instituto de F\'{i}sica, Universidad Nacional Autónoma de México, Ciudad de Mexico, Mexico }

\author[0000-0001-8644-4734]{M.~Schneider}
\affiliation{Department of Physics, University of Maryland, College Park, MD, USA }

\author[0000-0002-8999-9249]{H.~Schoorlemmer}
\affiliation{Max-Planck Institute for Nuclear Physics, D-69117 Heidelberg, Germany}

\author{J.~Serna-Franco}
\affiliation{Instituto de F\'{i}sica, Universidad Nacional Autónoma de México, Ciudad de Mexico, Mexico }

\author[0000-0002-1012-0431]{A.J.~Smith}
\affiliation{Department of Physics, University of Maryland, College Park, MD, USA }

\author[0000-0002-1492-0380]{R.W.~Springer}
\affiliation{Department of Physics and Astronomy, University of Utah, Salt Lake City, UT, USA }

\author[0000-0002-8516-6469]{P.~Surajbali}
\affiliation{Max-Planck Institute for Nuclear Physics, D-69117 Heidelberg, Germany}

\author[0000-0001-9725-1479]{K.~Tollefson}
\affiliation{Department of Physics and Astronomy, Michigan State University, East Lansing, MI, USA }

\author[0000-0002-1689-3945]{I.~Torres}
\affiliation{Instituto Nacional de Astrof\'{i}sica, Óptica y Electrónica, Puebla, Mexico }

\author{R.~Torres-Escobedo}
\affiliation{Departamento de F\'{i}sica, Centro Universitario de Ciencias Exactase Ingenierias, Universidad de Guadalajara, Guadalajara, Mexico }

\author[0000-0003-1068-6707]{R.~Turner}
\affiliation{Department of Physics, Michigan Technological University, Houghton, MI, USA}

\author[0000-0002-2748-2527]{F.~Ureña-Mena}
\affiliation{Instituto Nacional de Astrof\'{i}sica, Óptica y Electrónica, Puebla, Mexico }

\author[0000-0001-6876-2800]{L.~Villaseñor}
\affiliation{Facultad de Ciencias F\'{i}sico Matemáticas, Benemérita Universidad Autónoma de Puebla, Puebla, Mexico }

\author{T.~Weisgarber}
\affiliation{Department of Physics, University of Wisconsin-Madison, Madison, WI, USA}

\author{E.~Willox}
\affiliation{Department of Physics, University of Maryland, College Park, MD, USA }

\author[0000-0003-0513-3841]{H.~Zhou}
\affiliation{Tsung-Dao Lee Institute \&{} School of Physics and Astronomy, Shanghai Jiao Tong University, Shanghai, People's Republic of China }

\collaboration{91}{(HAWC collaboration)}

%% Note that the \and command from previous versions of AASTeX is now
%% depreciated in this version as it is no longer necessary. AASTeX 
%% automatically takes care of all commas and "and"s between authors names.

%% AASTeX 6.3 has the new \collaboration and \nocollaboration commands to
%% provide the collaboration status of a group of authors. These commands 
%% can be used either before or after the list of corresponding authors. The
%% argument for \collaboration is the collaboration identifier. Authors are
%% encouraged to surround collaboration identifiers with ()s. The 
%% \nocollaboration command takes no argument and exists to indicate that
%% the nearby authors are not part of surrounding collaborations.

%% Mark off the abstract in the ``abstract'' environment. 

\begin{abstract}

   The High Altitude Water Cherenkov (HAWC) observatory and the High Energy Stereoscopic System (H.E.S.S.) are two leading instruments in the ground-based very-high-energy $\gamma$-ray domain. HAWC employs the water Cherenkov detection (WCD) technique, while H.E.S.S. is an array of Imaging Atmospheric Cherenkov Telescopes (IACTs). The two facilities  therefore differ in multiple aspects, including their observation strategy, the size of their field of view and their angular resolution, leading to different analysis approaches. Until now, it has been unclear if the results of observations by both types of instruments are consistent: several of the recently discovered HAWC sources have been followed up by IACTs, resulting in a confirmed detection only in a minority of cases. 
   With this paper, we go further and try to resolve the tensions between previous results by performing a new analysis of the H.E.S.S. Galactic plane survey data, applying an analysis technique comparable between H.E.S.S. and HAWC. 
   Events above 1 TeV are selected for both datasets, the point spread function of H.E.S.S. is broadened to approach that of HAWC, and a similar background estimation method is used. 
   This is the first detailed comparison of the Galactic plane observed by both instruments. H.E.S.S. can confirm the $\gamma$-ray emission of four HAWC sources among seven previously undetected by IACTs, while the three others have measured fluxes below the sensitivity of the H.E.S.S. dataset. Remaining differences in the overall $\gamma$-ray flux can be explained by the systematic uncertainties.
   Therefore, we confirm a consistent view of the $\gamma$-ray sky between WCD and IACT techniques.

\end{abstract}

%% Keywords should appear after the \end{abstract} command. 
%% See the online documentation for the full list of available subject
%% keywords and the rules for their use.

%% From the front matter, we move on to the body of the paper.
%% Sections are demarcated by \section and \subsection, respectively.
%% Observe the use of the LaTeX \label
%% command after the \subsection to give a symbolic KEY to the
%% subsection for cross-referencing in a \ref command.
%% You can use LaTeX's \ref and \label commands to keep track of
%% cross-references to sections, equations, tables, and figures.
%% That way, if you change the order of any elements, LaTeX will
%% automatically renumber them.
%%
%% We recommend that authors also use the natbib \citep
%% and \citet commands to identify citations.  The citations are
%% tied to the reference list via symbolic KEYs. The KEY corresponds
%% to the KEY in the \bibitem in the reference list below. 

\section{Introduction}
The High Energy Stereoscopic System (H.E.S.S.) telescope array and the High Altitude Water Cherenkov (HAWC) observatory are examples of instruments using the two main techniques developed in very high energy (VHE) $\gamma$-ray astronomy. 
H.E.S.S. is an array of five Imaging Atmospheric Cherenkov Telescopes (IACTs) that image the Cherenkov light produced by the atmospheric air showers. H.E.S.S. is located in Namibia at a latitude of approximately 23\textdegree \ south, and at an altitude of 1800 m. Its collection area is of the order of $10^5$ m$^2$ above 1~TeV. H.E.S.S. has a field of view of 5\textdegree \ in diameter, with a $\gamma$-ray sensitivity roughly uniform for the innermost 2\textdegree \ that gradually drops toward the edges. 
Its point-spread function (PSF) is better than 0.1\textdegree \ above 1~TeV for zenith angles less than 30\textdegree~\citep{HESS_crab}. 
H.E.S.S. is a pointing instrument observing only at night. 
Including CT5, the biggest telescope at the center of the array, it is sensitive in the energy range 30~GeV~-~100~TeV, and its energy resolution is better than 15\% above 1~TeV. 
H.E.S.S. surveyed the Galactic plane from 2004 to 2013, producing the H.E.S.S. Galactic Plane Survey (HGPS;~\cite{HGPS}) based on 2700 hours of selected data, with 78 detected $\gamma$-ray sources. 
The survey covers the inner Galactic plane from 250\textdegree\ to 65\textdegree\  in longitude and from $-3$\textdegree\ to $+3$\textdegree\ in latitude, extending up to $-5$\textdegree \ and $+5$\textdegree \ in some regions. This observation program was carried out by the four middle-size telescopes, known as H.E.S.S. phase I. The energy threshold for the hard cuts used in making emission maps is, for most ranges of Galactic longitude, between 400 and 700~GeV (see Figure~2 of~\citet{HGPS}). The achieved sensitivity is better than 2\% of the Crab flux, assuming a point source with a spectral index of $-2.3$, and improves to 0.5\% in some regions with high observation time, like in the vicinity of the Galactic Centre.

The second observatory, HAWC, is an array of 300 Water Cherenkov Detectors (WCDs) located at a latitude of about 19\textdegree \ north in Mexico, at an altitude of 4100 m. The collection area of HAWC is of the order of the physical area of the detector, $\sim$10$^4$~m$^2$ above 500 GeV for a source at the Crab declination. HAWC detects the Cherenkov light produced in the water by the individual charged particles of air showers. Since HAWC is a survey instrument, it continuously monitors the sky above it. It has an instantaneous field of view of $\sim$2~sr and can observe two-thirds of the sky from $\sim -20$\textdegree \ to $\sim +60$\textdegree \ in declination, every day. The achieved sensitivity is $\sim$2\% of the Crab flux for sources with declination 0\textdegree~<~$\delta$~<~40\textdegree \ assuming a spectral index of $-2.5$, and degrades to 8\% of the Crab flux for sources at the edge of HAWC observable sky ($\delta=-20$\textdegree \ and +60\textdegree).  HAWC is sensitive in the energy range between $\sim$300~GeV to above 100~TeV, but is optimized to $\gamma$ rays with a primary energy above $\sim$1~TeV. The detected events are allocated to one of the nine analysis bins, depending on the fraction of the array that triggered, from analysis bin 1 gathering events triggering $7-10$\% of the array, to analysis bin 9 for events hitting $84-100$\% of the array. This binning is linked to the energy of the incoming particle, since events triggering only a small fraction of the array are likely to have lower energy than events triggering most of the array. However, there is some degeneracy and each analysis bin spans roughly one order of magnitude of energy, making HAWC's energy resolution quite poor as shown by~\citet{HAWC_crab}, Figure 3. 
The PSF of HAWC is $\sim$1$^{\circ}$ for the first analysis bin and improves to $\sim$0.2$^{\circ}$ for the ninth analysis bin, as depicted in Figure 9 of~\citet{HAWC_crab}. 

The HAWC Collaboration published its 2HWC catalog based on 507 days of data~\citep{HAWC_catalog} as the result of the first source search performed with the complete HAWC detector. More recently, using three times more data, the HAWC Collaboration published the 3HWC catalog~\citep{3HWC_catalog}. In this search, 65 sources have been reported, among them 20 detected for the first time at teraelectronvolt energies. 
The newly detected unassociated sources were subsequently studied by IACT collaborations who looked for counterparts. The VERITAS Collaboration (Very Energetic Radiation Imaging Telescope Array System) for example, looked at 14 2HWC sources and confirmed only one~\citep{HAWC_Fermi_VERITAS}. Archival data taken with the Major Atmospheric Gamma Imaging Cherenkov (MAGIC) telescopes were analyzed again to look for three unassociated 2HWC sources, without detecting any significant emission~\citep{HAWC_Fermi_MAGIC}. One explanation would be that these sources might be relatively extended, with a low surface brightness and possibly a hard spectrum. IACTs, with their limited field of view and their very good angular resolution, may not be sensitive to angularly extended objects with low surface brightness, using their standard analysis. 
We show in this paper that a dedicated analysis of IACT data is required to detect the extended and faint HAWC sources.

IACTs and WCDs have already demonstrated that they complement each other, with their overlapping observable sky and similar energy range yet different field of view and angular resolution. One example is the source MGRO~J1908+06, first detected by the MILAGRO observatory~\citep{MGROJ1908+06}, the predecessor of HAWC, and confirmed by H.E.S.S.~\citep{HESSJ1908+063}. 
In this paper we go further in exploiting their complementarity. Using a new analysis for the H.E.S.S. data, we present a comparison of the part of the Galactic plane common to H.E.S.S. and HAWC. 
After describing the data sets in section~\ref{data_set}, section~\ref{analysis} presents the new analysis method that aims at producing a H.E.S.S. Galactic plane map in the most similar way to HAWC. The results are presented in section~\ref{results_and_discussion} and the conclusion in section~\ref{conclusion}.

%__________________________________________________________________

\section{Data set}
\label{data_set}
\subsection{HAWC data}
The analysis presented here uses 1523 days of data, taken between 2014 November 26 and 2019 June 3 and makes use of events falling in the analysis bins $4-9$. As defined in~\cite{HAWC_crab}, these are events that triggered more than $\sim$25\% of the array. The main motivation for this selection is to have a reasonably good reconstruction, sufficient statistics, and a reasonable PSF. These cuts result in an energy threshold of approximately 1 TeV for a source at the Crab declination, increasing toward the central Galaxy. The PSF of the instrument is 0.4\textdegree \ for bin 4 and decreases to less than 0.2\textdegree \ for the highest bin.

\subsection{H.E.S.S. data}
\label{H.E.S.S. data}
The maps used to perform the H.E.S.S. analysis presented here have been made using the HGPS data set~\citep{HGPS} for Galactic longitude 10\textdegree~<~$\ell$~<~60\textdegree, reconstructed with the Image Pixel-wise fit for Atmospheric Cherenkov Telescopes (ImPACT) algorithm~\citep{ImPACT}. ImPACT uses a maximum likelihood approach to determine the best-fit shower parameters instead of using the traditional parametrization of the camera images~\citep{Hillas_1985} which was used in the published HGPS~\citep{HGPS}. Using ImPACT reconstruction, the PSF of H.E.S.S. is about 0.06\textdegree \ at 1 TeV for zenith angles smaller than 30\textdegree \ and its energy resolution is $\sim$10\% above 1~TeV. 
In this analysis, events above 1 TeV are selected to match the HAWC energy range. 

The results presented here have been verified with a different calibration, reconstruction, and gamma/hadron separation method using a semi-analytical description of the air shower~\citep{Modelpp}.

%__________________________________________________________________

\section{Analysis}
\label{analysis}

\begin{figure*}[h!]
   \centering
   \includegraphics[width=1\linewidth]{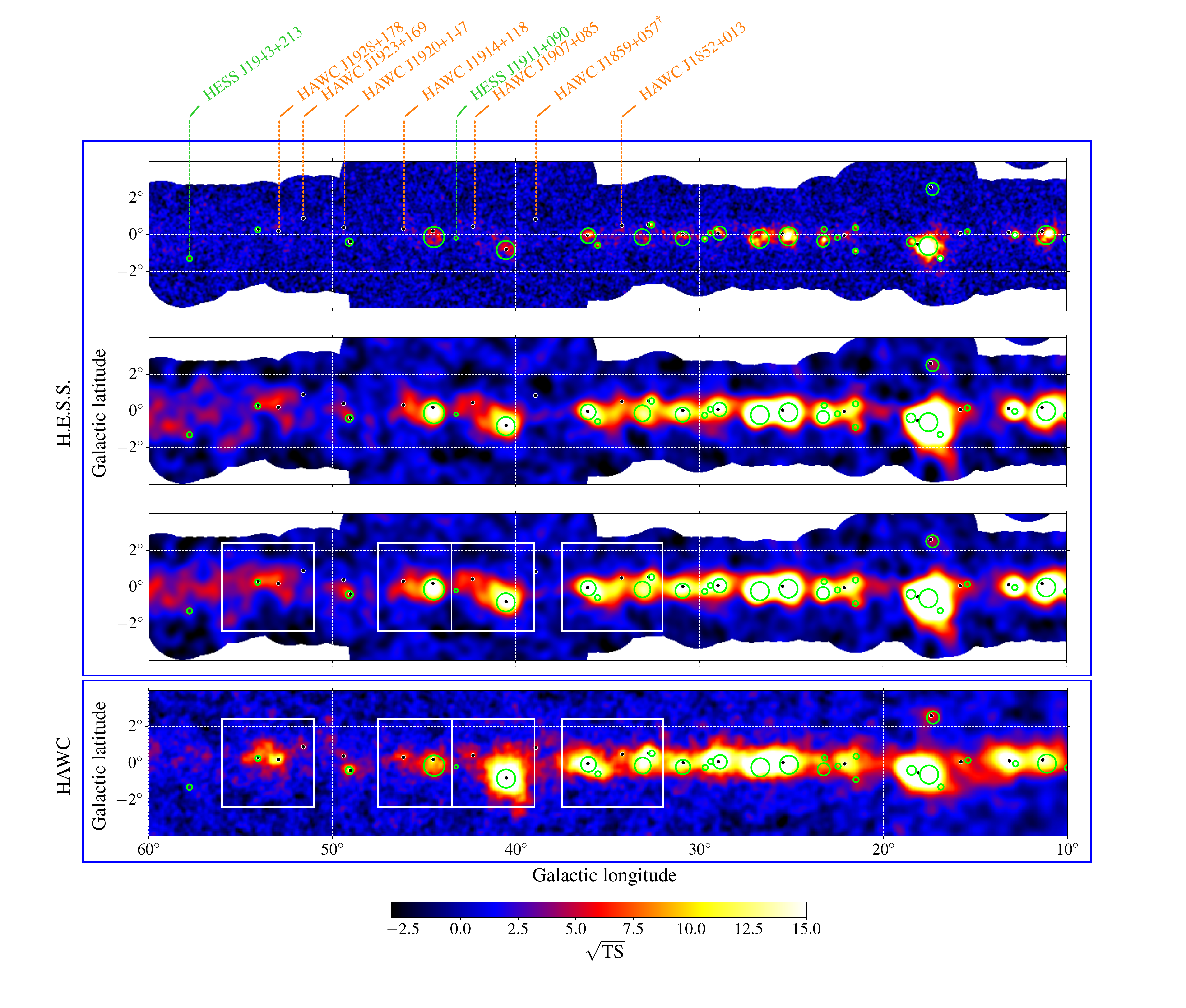}
    \caption{HAWC and H.E.S.S. Galactic plane maps. 
    The green circles are the 68\% containment of the HGPS sources and the black dots are the location of the HAWC sources. The regions in the white boxes are highlighted in Figure~\ref{new_sources}. The green dotted lines point to H.E.S.S. sources undetected by HAWC. The orange dotted lines point to HAWC sources previously undetected by H.E.S.S. (see section~\ref{new_hess_sources}). From top to bottom: 
     (1) H.E.S.S. Galactic plane map for E > 1 TeV, using ImPACT reconstruction and a Gaussian kernel of size 0.1\textdegree; the ring background method is applied on each observation run separately, with an adaptive radius. The standard exclusion regions around sources are used. 
     (2) Same H.E.S.S. data, using a Gaussian kernel of 0.4\textdegree. 
     (3) Same H.E.S.S. data, using the field-of-view background method and a Gaussian kernel of 0.4\textdegree. For the background normalization, in addition to the standard exclusion regions around sources, a 2\textdegree \ wide exclusion band covering the Galactic plane is used. 
     (4) HAWC Galactic plane map with 1523 days of data, using events in the analysis bins $4-9$.
    }
   \label{GPall}
\end{figure*}

For both data sets described in the previous section, a test statistic (TS) map is created with Gammapy~\citep{gammapy:2017} using a data map, a background map, and an exposure map. The TS is obtained by computing the likelihood ratio between the hypothesis of a source model, and a null hypothesis assuming no source but background only, for the test source centered on each pixel of the sky map. The source image through the instrument is assumed to have the shape of a Gaussian with a fixed angular extent $\sigma$. 
The flux normalization is the only free parameter.
According to Wilks theorem~\citep{Wilks}, the distribution of the TS under the null hypothesis follows asymptotically a $\chi^2$ distribution with one degree of freedom (see appendix \ref{mapsimu}).
The statistical significance is then equal to $\pm\sqrt{\mbox{TS}}$. Note that the significance mentioned thereafter is not corrected for trials.
The flux is obtained by dividing the excess that maximizes the likelihood by the $\gamma$-ray exposure.
The first row of Figure~\ref{GPall} shows the significance maps of the Galactic plane detected by H.E.S.S., above $\sim$1~TeV, made in a similar way to the published HGPS:  using a Gaussian of size $\sigma = 0.1$\textdegree \ and the ring background method (described in \cite{HGPS}). The background estimate is normalized in a ring region surrounding the sources, where detected sources or suspected $\gamma$-ray emission have been excluded. A spectral index of $-2.5$ is used, while it is $-2.3$ in the HGPS.

For the HAWC map, due to the large variation of the PSF with analysis bin, the TS map is produced by a maximum likelihood method over all analysis bins $4-9$. A simple power law with a spectral index of $-2.5$ is used as the spectral assumption. More details can be found in~\cite{HAWC_catalog}. A new catalog search has been performed on the HAWC data selected for this analysis, following the procedure described in~\citet{3HWC_catalog} and assuming sources that are point-like on the scale of HAWC angular resolution. The HAWC sources presented here, using the prefix HAWC at the beginning of their name, result from this new source search. They are listed in Table~\ref{tab:source1}, in appendix~\ref{HAWC_sources}, with their fitted position and their counterpart in the 3HWC catalog.

\begin{figure*}[ht!]
  \centering
  \includegraphics[width=0.54\linewidth]{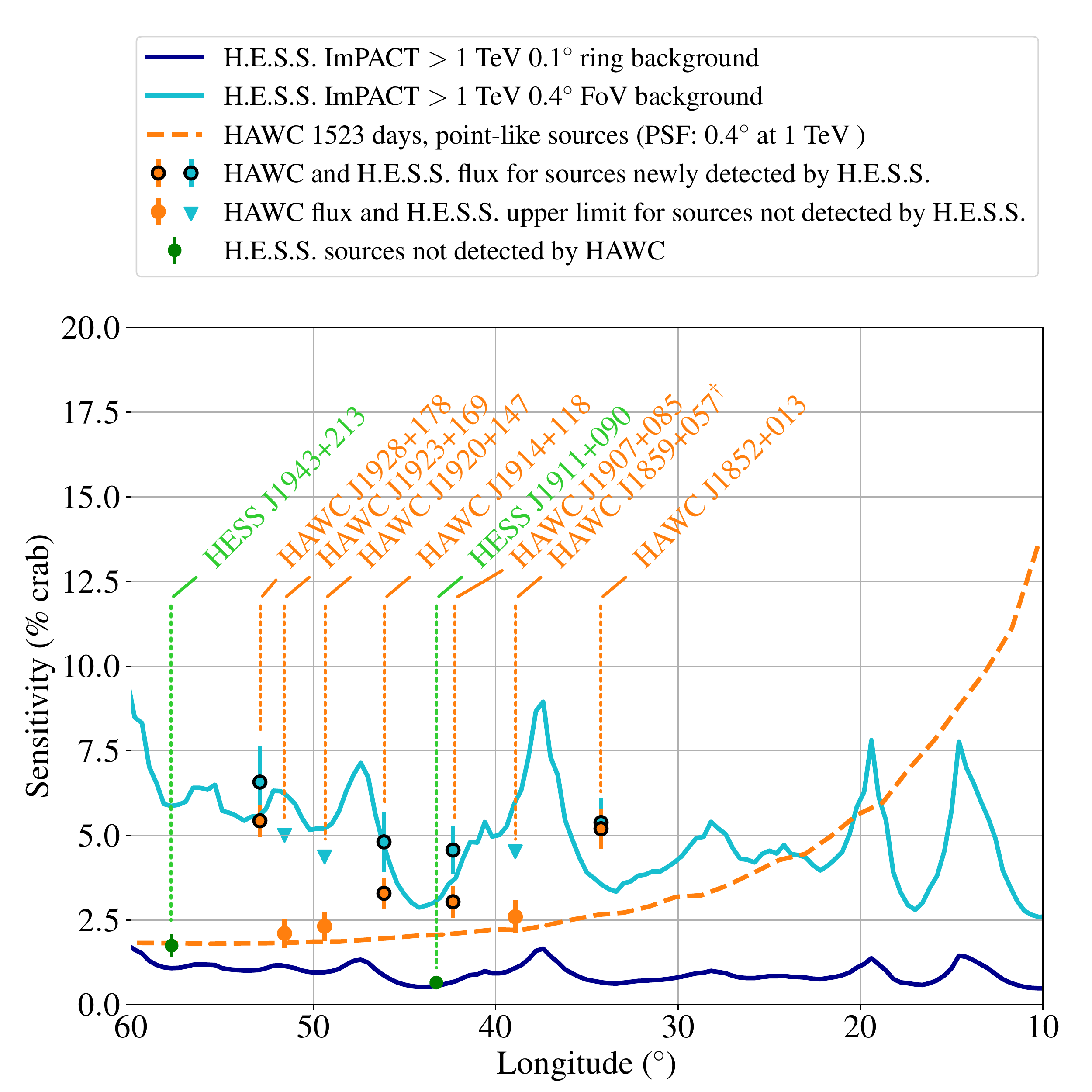} 
  \includegraphics[width=0.44\linewidth]{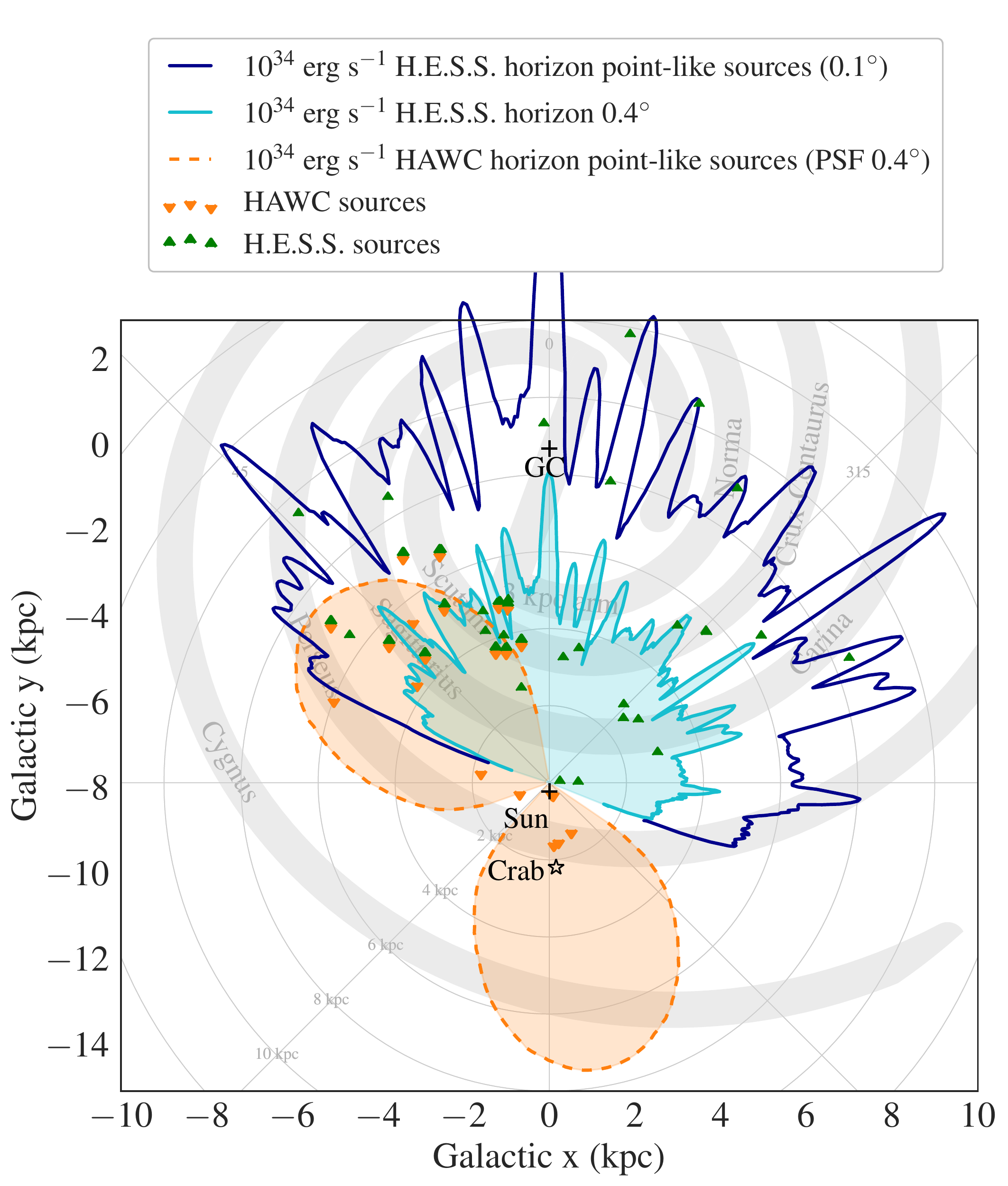} 
   \caption{Left: sensitivity curves at $b = 0^{\circ}$ corresponding to the first (H.E.S.S. point-like source analysis, dark blue curve), third (new H.E.S.S. extended source analysis, light blue curve), and fourth maps (HAWC, dashed orange) of Figure~\ref{GPall}. The orange and blue points represent the flux in Crab units of the seven HAWC sources that were previously undetected by H.E.S.S., obtained using the analysis presented here at the location of the HAWC sources. The black borders highlight the four HAWC sources newly detected by H.E.S.S., and they lie above the sensitivity curve corresponding to this analysis. The flux values can be found in Table~\ref{tab:source2} of Appendix~\ref{HAWC_sources} . The green points correspond to the flux of the two H.E.S.S. sources not detected by HAWC and they lie below the HAWC sensitivity curve. Note that 1 crab unit is equal to $2.26 \times 10^{-11}$~cm$^{-2}$~s$^{-1}$ for H.E.S.S. (from~\citet{HGPS}) and $2.24 \times 10^{-11}$~cm$^{-2}$~s$^{-1}$ for HAWC (from~\citet{3HWC_catalog}).
   Right: face-on view of the galaxy. The H.E.S.S. horizon at  $b = 0$\textdegree \ for a 5$\sigma$ detection of a point-like source is depicted in dark blue. The H.E.S.S. horizon corresponding to this analysis is depicted in light blue. The HAWC horizon is shown in dashed orange. The horizons are produced using the sensitivity curves from the left plot (plotted here for the longitude range corresponding to the comparison) for a source luminosity of 10$^{34}$~erg~s$^{− 1}$. The spiral arms of the galaxy~\citep{spiral_arms} are schematically drawn in gray. The green and orange triangles represent the sources detected by H.E.S.S. and HAWC, respectively for which a distance is known. The sources located further than the horizon must be brighter than 10$^{34}$~erg~s$^{− 1}$. Adapted from the ~\cite{HGPS}}
  \label{sensitivity_horizon}
\end{figure*}

In the spirit of making the Galactic plane maps from both instruments in the most similar way, a new H.E.S.S. TS map is computed using a Gaussian of $\sigma = 0.4$\textdegree, comparable to the size of the largest HAWC PSF for our dataset. It corresponds to the PSF of the analysis bin 4 where the most events are expected. This has as direct consequence to decrease the flux sensitivity as more background is included, which is visible in the Figure~\ref{sensitivity_horizon}. The corresponding significance map is displayed in the second row of Figure~\ref{GPall}. 
Finally, another background estimation method called the field-of-view background method~\citep{HESS_background} is used, as it is the closest one to the HAWC direct integration background method~\citep{HAWC_crab}. The field-of-view method uses as background the H.E.S.S. exposure map, \textit{i.e.} the expected events from cosmic-ray background, taking advantage of the whole field of view of H.E.S.S. and not only a small area contained in a ring around the source position.  
The shape of the background map is estimated from regions of the sky at similar zenith angle, excluding known VHE $\gamma$-ray sources. The map is normalized to the number of events outside the exclusion regions. 
However, note that using an average exposure assumes that its shape does not change with time and is not an instantaneous measurement as it does not fully account for variations in the instrument response over time, contrary to the HAWC direct integration method. Moreover, the required normalization step uses the number of events outside exclusion regions in the field of view, which has a limited size. 
The third map of Figure~\ref{GPall} shows the H.E.S.S. significance map produced using a Gaussian of 0.4\textdegree, similarly to the map in the second row, but using the field-of-view background method instead of the ring background method.
Furthermore, an additional exclusion region defined by a 2\textdegree \ wide strip centered on the Galactic plane (-1\textdegree~<~$b$~<~1\textdegree) is used, to avoid any contamination from the Galactic plane large scale $\gamma$-ray emission to the background estimation. 
This is the H.E.S.S. map produced with the most similar analysis method to the HAWC one, visible on the fourth row.

\section{Results and discussion}
\label{results_and_discussion}

\begin{figure*}[t!]
   \centering
   \includegraphics[width=1\linewidth]{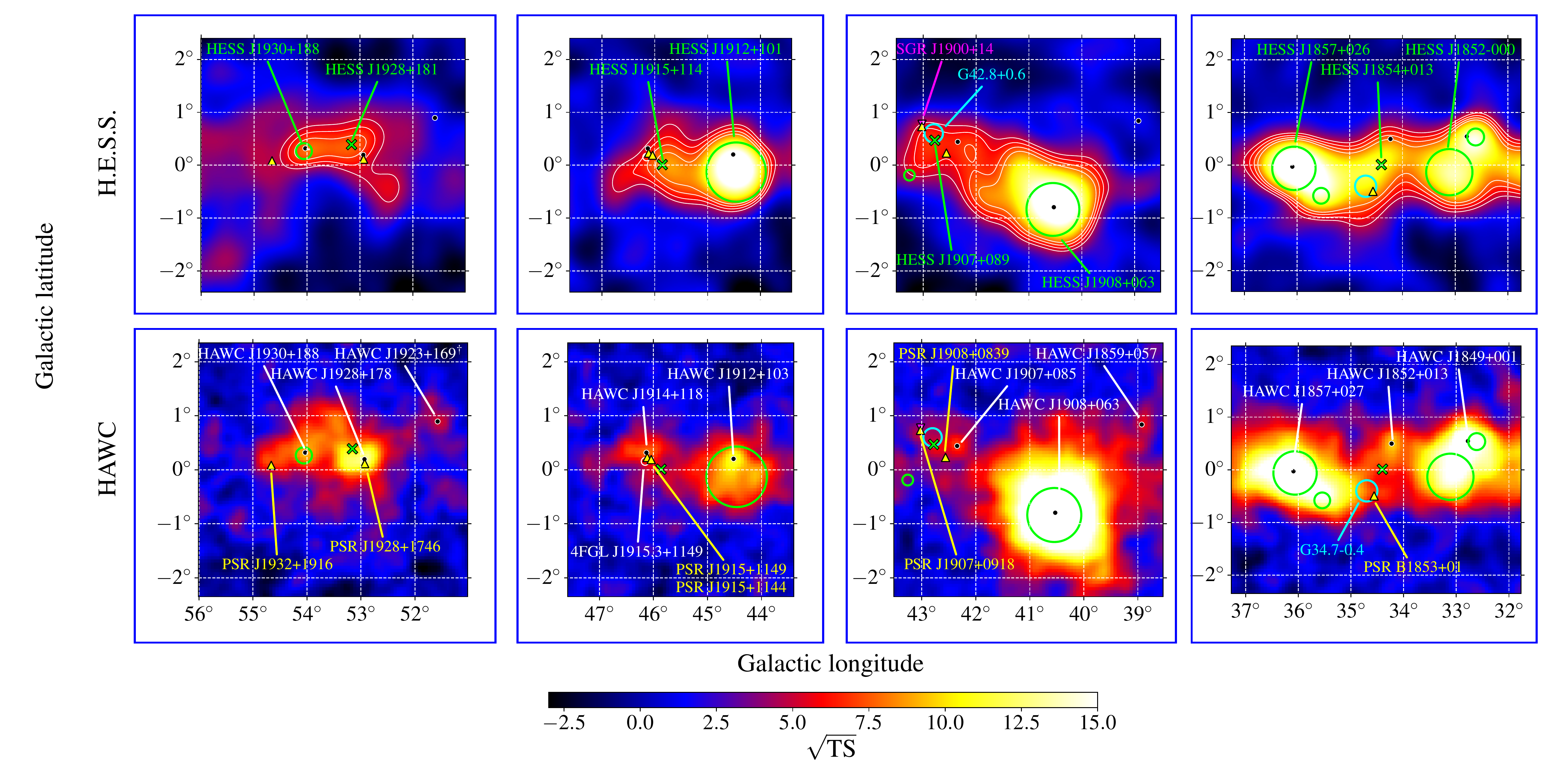}
    \caption{Regions in the white boxes of the third and fourth panels of Figure~\ref{GPall} showing the four HAWC sources newly undetected by H.E.S.S. using a HAWC-like map-making process. The white contours represent the 5$\sigma$, 6$\sigma$, 7$\sigma$ and 8$\sigma$ significance levels. The green circles are the 68\% containment of the HGPS sources and the black dots are the location of the HAWC sources. The new H.E.S.S. sources are indicated with the green crosses. Some possible counterparts are highlighted using yellow triangles (pulsars), magenta triangle (magnetar) and cyan circle (SNR). Note that the HAWC sources are considered as point-like for HAWC. Their 1$\sigma$ statistic uncertainty is of the order of 0.1\textdegree.
    }
   \label{new_sources}
\end{figure*}

\subsection{HAWC and H.E.S.S. original maps}
The H.E.S.S. significance map in the first row of Figure~\ref{GPall}, produced in a traditional way with a Gaussian size $\sigma$ = 0.1\textdegree \ and the ring background method, shows some important differences compared to the HAWC significance map, in the last row of the same Figure. The corresponding point-like sensitivity curves at $b = 0$\textdegree \ are displayed in the left panel of Figure~\ref{sensitivity_horizon} in dark blue and dashed orange, respectively, for the part of the Galactic plane studied here. The right panel of Figure~\ref{sensitivity_horizon} shows the detection horizon corresponding to the sensitivity curves plotted in the left panel of the same Figure, for a source luminosity of $10^{34}$~erg~s$^{−1}$. It illustrates the difference in exposure between both instruments: H.E.S.S. has an exposure time that varies significantly from one region to another, while the HAWC exposure changes very smoothly across the sky. It also illustrates the part of the sky covered by HAWC and by H.E.S.S. for the HGPS. 

Among the 29 sources detected by H.E.S.S. in the selected region of the HGPS, HAWC detected significant emission in at the location of 27 of them, even though only 15 sources are actually resolved. The two HGPS sources that do not show significant emission in the HAWC map are: HESS~J1943+213, which is likely to be an extragalactic source seen in the Galactic plane~\citep{HGPS} and HESS~J1911+090 associated with the supernova remnant (SNR) W49B. This is expected because the flux of these sources lies below the HAWC sensitivity curve, as shown in Figure~\ref{sensitivity_horizon}. On the other hand, HAWC detected 22 sources and only 15 have a counterpart in the HGPS catalog. Hence, seven sources are detected only by HAWC. 

There are several possible explanations for the different number of sources detected by HAWC and H.E.S.S.:
\begin{itemize}

    \item The resolution of H.E.S.S. at 1~TeV is several times better than the resolution of HAWC for analysis bin 4. Since its PSF is wider, HAWC cannot resolve all the H.E.S.S. sources and several H.E.S.S. sources are seen as only one by HAWC.

    \item With its 0.1\textdegree \ kernel and the ring background subtraction, the HGPS analysis is optimized for sources that are point-like or have a modest extension. Extended sources with low surface brightness are challenging for this kind of analysis. The variation of sensitivity as a function of the source extension is illustrated by ~\citet{HGPS}, Figure~13. The seven HAWC sources undetected by H.E.S.S. lie above the H.E.S.S. sensitivity curve (dark blue curve in Figure~\ref{sensitivity_horizon}), which means that if their image was a gaussian of the order of 0.1\textdegree \ in size, H.E.S.S. would have detected them. 
    Therefore, they likely have a larger extension. 
    
    \item The H.E.S.S. sensitivity peaks at lower energy than HAWC, as illustrated by ~\citet{HAWC_crab}, Figure~15: above $\sim$10 TeV, gamma/hadron separation in HAWC becomes very efficient and detection is not limited by the background anymore but by the number of arriving photons. Hence, continuous observing instruments like HAWC take advantage over short exposure instruments like H.E.S.S.

\end{itemize}{}

\subsection{HAWC sources newly detected by H.E.S.S.}
\label{new_hess_sources}
Using the new analysis presented in section~\ref{analysis}, a new significance map is produced, shown in the third row of Figure~\ref{GPall}. Its corresponding sensitivity curve is plotted in light blue in the left panel of Figure~\ref{sensitivity_horizon}. Statistically significant emission $>5\sigma$ is now visible next to four of the previously undetected HAWC sources: HAWC~J1928+178, HAWC~J1914+118, HAWC~J1907+085 and HAWC~J1852+013. These sources are visible in the four panels of Figure~\ref{new_sources}, with their possible counterparts. White contours indicate the significance level. For each new H.E.S.S. source, a maximum of emission is identified in the flux map produced together with the TS map, to give a preliminary estimate of the location of the H.E.S.S. source.
In the case of the H.E.S.S. source corresponding to HAWC~J1852+013, the location of maximum flux is identified by increasing the energy threshold to 2 and 5 TeV in order to lower the contamination from the extended emission from the neighboring sources and possibly from large scale diffuse $\gamma$-ray emission. The four new H.E.S.S. sources are located 0.3\textdegree$-$ 0.5\textdegree \ away from the HAWC source.

\textbf{HESS~J1928+181} is the H.E.S.S. detection of the source of unknown origin HAWC J1928+178. It may be associated with the middle-aged pulsar (82 kyr)  PSR~J1928+1746~\citep{ICRC_J1928, MyThesis}, located 4.3 kpc away. Detected in radio wavelength~\citep{Discovery_PSRJ1928}, it has a spin-down power of $\dot{\mbox{E}}=1.6\times 10^{36}$~erg~s$^{-1}$. In that case, the $\gamma$-ray emission would come from inverse Compton (IC) scattering of e$^{\pm}$ from the associated pulsar wind nebula (PWN) with ambient photons. However, the lack of an X-ray counterpart, despite the 90 ks exposure of NuSTAR (the Nuclear Spectroscopic Telescope Array) and 20 ks of Chandra~(\citet{Chandra_10ks_J1928}, \citet{J1928_dark_accelerator}), may indicate the evolved state of the PWN. Alternatively, accelerated protons from the PWN may be interacting with a nearby molecular cloud and produce $\gamma$-ray emission via $\pi^0$ decay~\citep{MyThesis}. \cite{J1928_dark_accelerator} proposed a third scenario, involving a binary system where the variable X-ray source CXO~J192812.0+174712 would be a massive star with a pulsar companion. In that case, e$^{\pm}$ from the pulsar wind would be accelerated at the shock between the pulsar and the stellar wind of the massive star and produce $\gamma$ rays via synchrotron emission and IC scattering.

\textbf{HESS~J1915+115} is the H.E.S.S. counterpart of HAWC~J1914+118, an unidentified source without any obvious counterpart. Two pulsars detected in radio wavelength~\citep{Arecibo_pulsar_survey} are found in the vicinity: PSR~J1915+1149 and PSR~J1915+1144 located 0.09\textdegree \ and 0.16\textdegree \ away from the HAWC source. No spin-down power $\dot{\mbox{E}}$ has been reported. They are located at a distance of 14~kpc and 7.2~kpc respectively using \citet{ymw17}. Moreover, the \textit{Fermi}-LAT source 4FGL~J1915.3+1149 overlapps with PSR~J1915+1149 and the HAWC source within the location errors. 4FGL~J1915.3+1149 is associated with TXS~1913+115 and classified as an active galaxy of uncertain type, but the association probability is close to the threshold.

\textbf{HESS~J1907+089} is the H.E.S.S. detection of HAWC~J1907+085, and was formally reported in the HGPS as the hotspot HOTS~J1907+091. 
Two potential counterparts are spatially coincident: the magnetar SGR~1900+14~\citep{SGR1900} and the SNR~G42.8+0.6~\citep{SNRG42.8+0.6}. HAWC~J1907+085 is one of the new HAWC sources followed up by MAGIC and \textit{Fermi}-LAT~\citep{HAWC_Fermi_MAGIC} without being detected. A detailed study of SGR~1900+14 and its environment is presented in \cite{SGR1900+14}, including a model of the observed high energy (E > 100 MeV; 4FGL~J1908.6+0915e) and VHE $\gamma$-ray emission~(E~>~100~GeV; HOTS~J1907+091 and 3HWC~J1907+085).

\textbf{HESS~J1854+013}, finally, is a source of significant $\gamma$-ray emission in the H.E.S.S. map near HAWC~J1852+013. Source confusion is very obvious in this region and affects the accuracy of deriving a clear peak position. Note the presence in this region, $\sim$1 degree away from HAWC~J1852+013, of the SNR~G34.7-0.4 (W44), detected in the radio, X-ray, and GeV $\gamma$-ray bands. W44 hosts the pulsar PSR~B1853+01, a 20~kyr pulsar with $\dot{\mbox{E}}=4.3\times 10^{35}$~erg~s$^{-1}$ located $\sim$3~kpc away. W44 has not been detected by H.E.S.S. during the HGPS, but H.E.S.S. reported significant emission with only one of their analysis chains at the location of W44, without claiming a detection~\citep{HESS_W44}.  

For the three remaining HAWC sources undetected in the HGPS, HAWC~J1923+169$^{\dag}$, HAWC~J1920+147, and HAWC~J1859+057*, no emission is seen in the H.E.S.S. map with the new analysis. These sources lie below the corresponding H.E.S.S. sensitivity curve (light blue curve in Figure~\ref{sensitivity_horizon}). Hence, they are not expected to be detected by H.E.S.S. even with this new analysis.

\subsection{Longitude and latitude profiles}
\label{profiles}

\begin{figure*}[h]
   \centering
   \includegraphics[width=0.98\linewidth]{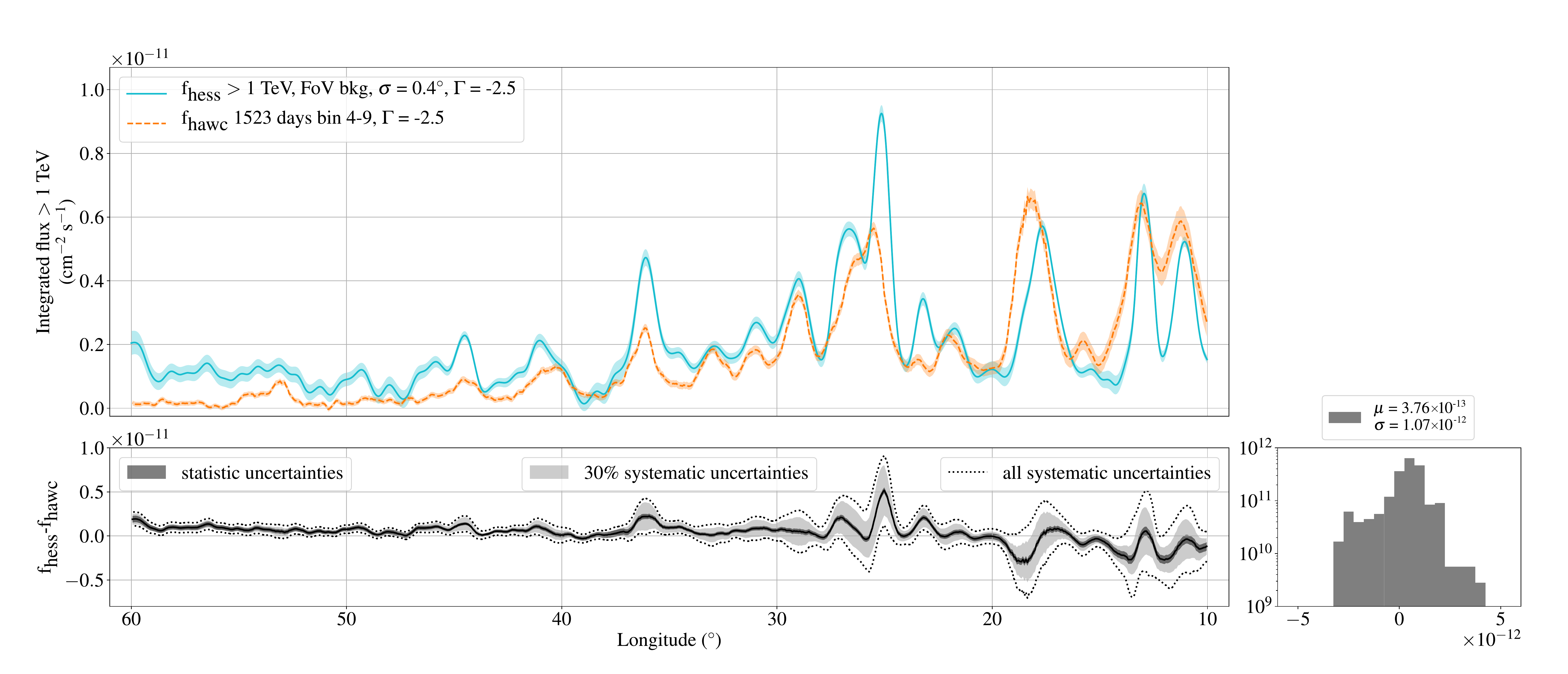}
    \caption{Longitude profiles of the integrated flux above 1~TeV at $b=0$\textdegree \ for the new H.E.S.S. map, using the 0.4\textdegree \ Gaussian and the field-of-view background method in cyan, and the HAWC map in dashed orange. The bottom panel shows the difference between the H.E.S.S. and HAWC flux and the corresponding histogram (mean and standard deviation given on top of the histogram). The systematic uncertainties are derived for each parameter used to produce the maps (listed in Appendix ~\ref{Systematics}) and added quadratically, are represented by the dotted envelop. More details on the derivation of the Systematic uncertainties can be found in Appendix~\ref{Systematics}.
    }
   \label{long_profile}
   \end{figure*}

Figure~\ref{long_profile} compares the longitude profiles at $b=0^{\circ}$ of the integrated flux above 1 TeV between the new H.E.S.S. map, using the 0.4\textdegree \ Gaussian the field-of-view background method and the large exclusion band, and the HAWC map obtained from the standard HAWC analysis. 
The overall shape of the profiles is in reasonable agreement. The difference between them and the corresponding histogram is plotted on the bottom panel. The integrated flux measured by H.E.S.S. is on average slightly higher than that of HAWC by $\mathrm{3.7 \times 10^{-13} \; cm^{-2} \; s^{-1}}$. Nevertheless, the flux values are compatible within the systematic uncertainties, described in Appendix~\ref{Systematics}, over most of the Galactic plane. 

\begin{figure*}[ht!]
   \centering
   \includegraphics[width=0.98\linewidth]{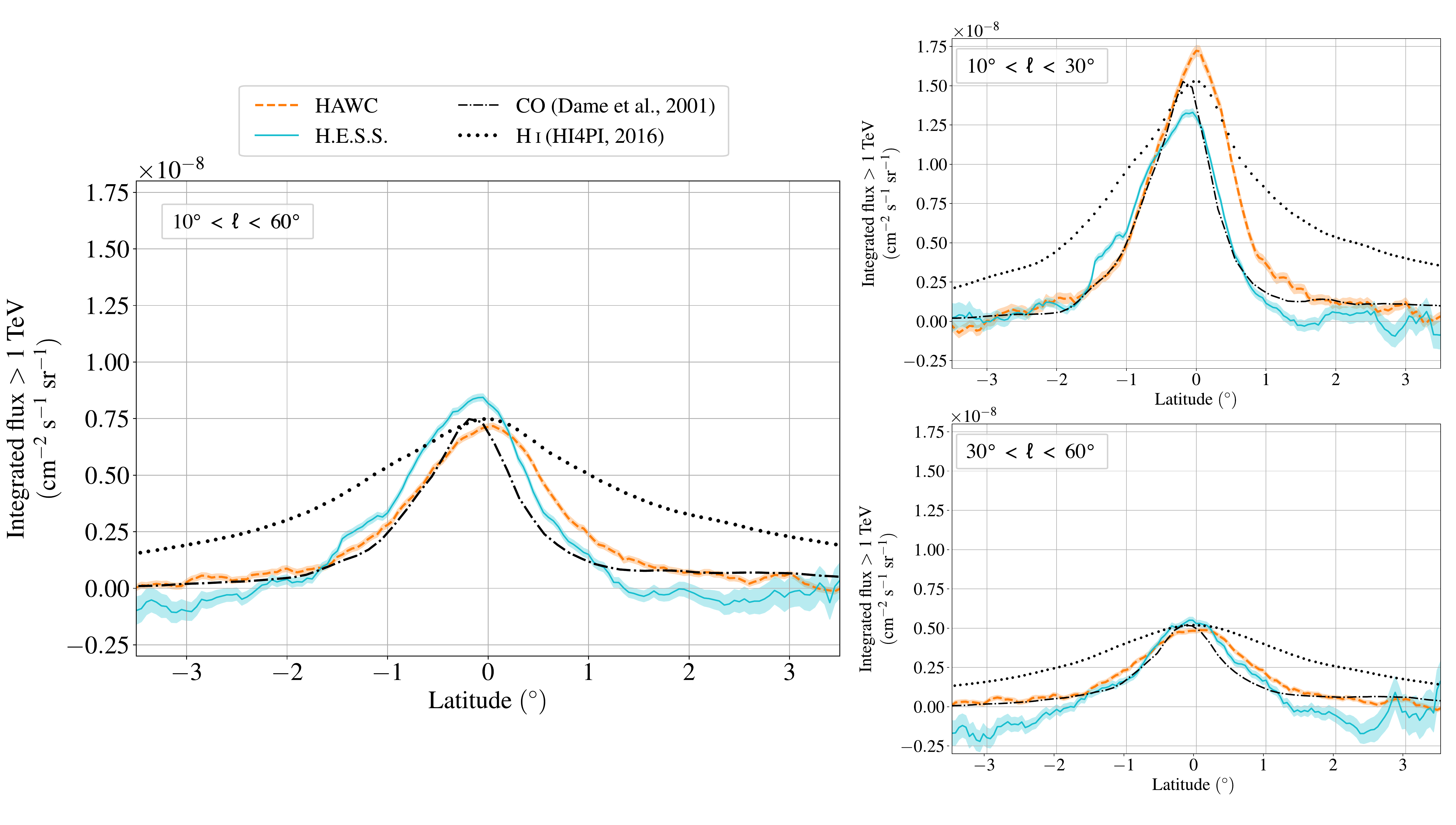}
    \caption{Profiles of the integrated flux above 1~TeV for H.E.S.S. and HAWC, along the latitude axis for different ranges in longitude ($\mathrm{10^{\circ} < \ell < 60^{\circ}}$, $\mathrm{10^{\circ} < \ell < 30^{\circ}}$ and $\mathrm{30^{\circ} < \ell < 60^{\circ}}$). The shaded bands represent the 1$\sigma$ statistic uncertainties. The systematic uncertainties are not shown here but from the study presented in the Appendix~\ref{Systematics} they can reach $\sim$50\%, depending on the longitude. Note that the sources have not been subtracted. The latitude profile extracted from the $^{12}$CO survey of \citet{COsurvey}, and the profile of the $\textrm{H}\scriptstyle\mathrm{~I}$ emission from the HI4PI survey~\citep{HI4PI} are also shown for the corresponding range in longitude, with an arbitrary normalization for display purposes.
    }
   \label{fig:lat_profile}
   \end{figure*}

Figure~\ref{fig:lat_profile} shows the latitude profiles of the integrated flux above 1~TeV for H.E.S.S. and HAWC. 
In this case, a sliding box of 0.4\textdegree \ width is defined with different lengths in longitude, as indicated in the different plots of the figure, that scans the Galactic plane in latitude between $-3.5$\textdegree \ and $+3.5$\textdegree.
In each of these boxes the surface brightness is derived by normalizing the average flux by the solid angle . 
The $^{12}$CO profile using data from~\citet{COsurvey} and the $\textrm{H}\scriptstyle\mathrm{~I}$ profile using the HI4PI survey from \citet{HI4PI} are also superimposed with an arbitrary normalization for display. 
The H.E.S.S. profile peaks at slightly negative latitude with a small asymmetry and seems to follow the $^{12}$CO profile, similar to what has been reported by~\citet{HGPS}, Figure~11, when comparing with the distribution of sources, or with~\citet{HESS_diffuse_emission}, Figure~2. 
The HAWC profile is more symmetric and centered around $b=0^{\circ}$, like the $\textrm{H}\scriptstyle\mathrm{~I}$ profile. 
The origin of this difference is not clear. Since the $\textrm{H}\scriptstyle\mathrm{~I}$  emission is very broad around the Galactic plane, the diffuse $\gamma$-ray emission coming from these high-latitude regions is likely removed in the H.E.S.S. analysis due to the background estimation method, but is conserved in the HAWC analysis because its very large field of view allows for a large exclusion region of $\pm 3$\textdegree \ around the Galactic plane. Hence, it could be the indication of an underlying diffuse emission component removed in the H.E.S.S. map. This may also explain that the H.E.S.S. profile tends to be slightly negative for $|b|>2$\textdegree. A more detailed study of the large scale diffuse emission is beyond the scope of this paper and will be the topic of a future paper. Another possible explanation for the shift in the H.E.S.S. profile with respect to $b = 0$\textdegree \ could be related to the HAWC bias in the source location reconstruction as a function of declination, reported by ~\citet{3HWC_catalog}, Figure~11. This would affect the overall shape toward positive latitude as it gets closer to the Galactic Centre.

Moreover, the latitude profile for $10^{\circ}~<~\ell~< 60^{\circ}$ shows that H.E.S.S. detects more flux than HAWC, while the profiles over smaller longitude bands (visible in the left plots of Figure~\ref{fig:lat_profile} and in Figure~\ref{fig:lat_profile_10deg}, in Appendix~\ref{more_latitude_profile}) show that HAWC detects more flux than H.E.S.S. for $\ell < 40$\textdegree. This can be explained by the fact that the exposure time is taken into account: more weight is given to regions with more exposure time, and regions with $\ell < 40$\textdegree \ have a low exposure time for HAWC and a high exposure time for H.E.S.S.

\section{Conclusion}
\label{conclusion}

HAWC and H.E.S.S. are two complementary instruments using different analysis techniques. This study is the first attempt to reconcile their data and this analysis shows consistent results: constructing the maps by using comparable procedures, a common energy range and comparable angular resolution leads to two remarkably similar maps. Overall, HAWC and H.E.S.S. measure similar fluxes. 
Nonetheless, there are some remaining differences that are only partially understood.
The difference in sensitivity as a function of energy may play a role since HAWC is more sensitive than H.E.S.S. at the highest energies, it will be more sensitive to sources with a hard spectrum. 
In the part of the Galactic plane common to H.E.S.S. and HAWC, four HAWC sources previously undetected by H.E.S.S. show significant emission above the detection level of 5$\sigma$, using the analysis presented here. The main ingredients leading to this result are the use of a large Gaussian smearing factor, followed by the field-of-view background estimation method together with a large exclusion band to avoid contamination from the Galactic plane.
It is also an excellent demonstration of the capability for an IACT to be sensitive to extended sources with low surface brightness emission. 
Each of these sources will require a dedicated analysis in order to study them in more detail and attempt to identify the possible origin of the observed $\gamma$-ray emission.  
Future work would benefit from combining the data from both instruments in a joint analysis, which will be supported by ongoing efforts toward multi-instrument open source tools like Gammapy~\citep{gammapy:2017} or the Multi-Mission Maximum Likelihood (3ML) framework~\citep{3ML}, together with a more realistic model of the exposure map as presented in \citet{Lars_bkg_model} for the background estimation.
The number of new extended $\gamma$-ray emission regions is expected to increase further by applying this new analysis to the southern part of the HGPS, which is out of HAWC's reach. 
The future observatories SWGO (Southern Wide-field Gamma-ray Observatory~\citep{SWGO_ICRC2019}) and  CTA (Cherenkov Telescope Array;~\citet{CTAbook}) can take advantage of the complementarity of the two detection techniques illustrated here. 
SWGO will observe the southern hemisphere with an improved sensitivity and angular resolution in comparison with HAWC. Combination of SWGO data with CTA data and their better control over systematic errors will allow broadband spectrum analysis and provide a deeper understanding of the sources.
 
%\acknowledgments
\section*{Acknowledgments}
\small
We acknowledge the support from the US National Science Foundation (NSF); the US Department of Energy Office of High-Energy Physics; the Laboratory Directed Research and Development (LDRD) program of Los Alamos National Laboratory; Consejo Nacional de Ciencia y Tecnolog\'ia (CONACyT), M\'exico, grants 271051, 232656, 260378, 179588, 254964, 258865, 243290, 132197, A1-S-46288, and A1-S-22784, c\'atedras 873, 1563, 341, and 323, Red HAWC, M\'exico; DGAPA-UNAM grants IG101320, IN111315, IN111716-3, IN111419, IA102019, and IN112218; VIEP-BUAP; PIFI 2012, 2013, PROFOCIE 2014, 2015; the University of Wisconsin Alumni Research Foundation; the Institute of Geophysics, Planetary Physics, and Signatures at Los Alamos National Laboratory; Polish Science Centre grant, DEC-2017/27/B/ST9/02272; Coordinaci\'on de la Investigaci\'on Cient\'ifica de la Universidad Michoacana; Royal Society - Newton Advanced Fellowship 180385; Generalitat Valenciana, grant CIDEGENT/2018/034; Chulalongkorn University’s CUniverse (CUAASC) grant; Coordinaci\'on General Acad\'emica e Innovaci\'on (CGAI-UdeG) and PRODEP-SEP through UDG-CA-499; Institute of Cosmic Ray Research (ICRR), University of Tokyo. Thanks to Scott Delay, Luciano D\'iaz and Eduardo Murrieta for technical support.

The support of the Namibian authorities and of the University of Namibia in facilitating the construction and operation of H.E.S.S. is gratefully acknowledged, as is the support by the German Ministry for Education and Research (BMBF), the Max Planck Society, the German Research Foundation (DFG), the Helmholtz Association, the Alexander von Humboldt Foundation, the French Ministry of Higher Education, Research and Innovation, the Centre National de la Recherche Scientifique (CNRS/IN2P3 and CNRS/INSU), the Commissariat à l’énergie atomique et aux énergies alternatives (CEA), the U.K. Science and Technology Facilities Council (STFC), the Knut and Alice Wallenberg Foundation, the National Science Centre, Poland grant No. 2016/22/M/ST9/00382, the South African Department of Science and Technology and National Research Foundation, the University of Namibia, the National Commission on Research, Science \& Technology of Namibia (NCRST), the Austrian Federal Ministry of Education, Science and Research and the Austrian Science Fund (FWF), the Australian Research Council (ARC), the Japan Society for the Promotion of Science, and the University of Amsterdam. We appreciate the excellent work of the technical support staff in Berlin, Zeuthen, Heidelberg, Palaiseau, Paris, Saclay, Tübingen and Namibia in the construction and operation of the equipment. This work benefited from services provided by the H.E.S.S. Virtual Organisation, supported by the national resource providers of the EGI Federation. 

%% To help institutions obtain information on the effectiveness of their 
%% telescopes the AAS Journals has created a group of keywords for telescope 
%% facilities.
%
%% Following the acknowledgments section, use the following syntax and the
%% \facility{} or \facilities{} macros to list the keywords of facilities used 
%% in the research for the paper.  Each keyword is check against the master 
%% list during copy editing.  Individual instruments can be provided in 
%% parentheses, after the keyword, but they are not verified.

\facility{HAWC (https://www.hawc-observatory.org/), H.E.S.S. (https://www.mpi-hd.mpg.de/hfm/HESS/)}

%% Similar to \facility{}, there is the optional \software command to allow 
%% authors a place to specify which programs were used during the creation of 
%% the manuscript. Authors should list each code and include either a
%% citation or url to the code inside ()s when available.

\software{Astropy (\cite{astropy:2013}, \cite{astropy:2018}),  
          Gammapy-0.13 and Gammapy-0.17 (\cite{gammapy:2017}, \cite{gammapy:2019}) }

%% Appendix material should be preceded with a single \appendix command.
%% There should be a \section command for each appendix. Mark appendix
%% subsections with the same markup you use in the main body of the paper.

%% Each Appendix (indicated with \section) will be lettered A, B, C, etc.
%% The equation counter will reset when it encounters the \appendix
%% command and will number appendix equations (A1), (A2), etc. The
%% Figure and Table counter will not reset.

\newpage

\appendix
\section{HAWC sources}
\label{HAWC_sources}

The HAWC sources in this analysis are the result of the HAWC source search performed on the latest HAWC data using 1523 days and events falling in the analysis bins $4-9$ only. According to the definition given in \cite{HAWC_crab}, they correspond to events triggering more than $\sim$25\% of the array that would have an energy greater than approximately 1 TeV for a source at the Crab declination. The procedure is the same as the one used for the 3HWC catalog~\citep{3HWC_catalog}. 
Table~\ref{tab:source1} shows the resulting list of sources with their position and the association with the nearby TeVCat\footnote{http://tevcat.uchicago.edu/}~\citep{TeVCat} sources within 1\textdegree \ of the HAWC source. 
Table~\ref{tab:source2} gives the integrated flux above 1 TeV from the HAWC flux map and the H.E.S.S. flux map using the new analysis at the location of the HAWC source reported in Table~\ref{tab:source1}. 
The sources highlighted in black bold are found both in the HAWC search and in the HGPS catalog. The sources highlighted in bold orange are the HAWC sources discussed in this analysis having no H.E.S.S. counterpart in the HGPS. 
The sources with a dagger (\dag) are called \textit{secondary sources}. They correspond to local maxima that have reached the detection threshold but not the separation criteria based on the TS gap between two neighbor sources, as defined in the 3HWC catalog~\citep{3HWC_catalog}.

\begin{table*}[!ht]
 \centering
 \caption{HAWC source list with their position. The position uncertainty reported here is statistical only. The systematic uncertainty varies with declination and is discussed in more details in Appendix~\ref{Systematics}. For each source, the nearby known TeV sources within 1\textdegree \ of the HAWC source, listed in the TeVCat~\citep{TeVCat} is also given. } 
 \label{tab:source1}.
 \begin{tabular}{c c c c c c c c}
  \hline
  \hline
  Name           &  TS    & RA            & Dec           & $\ell$             & $b$             & 1$\sigma$ stat. unc. & TeVCat Source     \\
                 &        & (\textdegree) & (\textdegree) & (\textdegree) & (\textdegree) & (\textdegree)        &                   \\
  
  \hline
 \textbf{HAWC~J1809-190}                 &   222.8 & 272.46 & -19.04 &  11.33 &   0.18 &  0.11  & 3HWC~J1809-190, \textbf{HESS J1809-193}\\
 \textbf{HAWC~J1813-125}                 &    42.5 & 273.30 & -12.56 &  17.40 &   2.59 &  0.15  & 3HWC~J1813-125, \textbf{HESS J1813-126}\\
 \textbf{HAWC~J1813-174}                 &   354.1 & 273.43 & -17.47 &  13.15 &   0.13 &  0.06  & 3HWC J1813-174, \textbf{HESS J1813-178}\\
 \textbf{HAWC~J1819-151}                 &    69.4 & 274.79 & -15.17 &  15.79 &   0.07 &  0.14  & 3HWC J1819-150$^{\dag}$, \textbf{HESS J1818-154}\\
 \textbf{HAWC~J1825-134}                 &  1856.6 & 276.46 & -13.40 &  18.12 &  -0.53 &  0.06  & 3HWC J1825-134, \textbf{HESS J1825-137}\\
 \textbf{HAWC~J1831-096}                 &   176.4 & 277.91 &  -9.63 &  22.11 &  -0.03 &  0.11  & 3HWC J1831-095, \textbf{HESS J1831-098}\\
 \textbf{HAWC~J1837-066}                 &  1111.9 & 279.40 &  -6.62 &  25.47 &   0.04 &  0.06  & 3HWC J1837-066, \textbf{HESS J1837-069}\\
 \textbf{HAWC~J1843-034}                 &   705.0 & 280.99 &  -3.47 &  28.99 &   0.08 &  0.06  & 3HWC J1843-034, \textbf{HESS J1843-033}\\
 \textbf{HAWC~J1847-017}                 &   226.8 & 281.91 &  -1.79 &  30.90 &   0.03 &  0.09  & 3HWC~J1847-017, \textbf{HESS J1848-018} \\
 \textbf{HAWC~J1849+001}                 &   372.4 & 282.31 &   0.11 &  32.78 &   0.54 &  0.06  & 3HWC~J1849+001, \textbf{HESS J1849-000}, \\
                                         &         &        &        &        &        &        & IGR J18490-0000 \\
 \color{orange} \textbf{HAWC~J1852+013}  &   93.7  & 283.01 &   1.38 &  34.23 &   0.50 &  0.10  & 3HWC J1852+013$^{\dag}$ \\
 \textbf{HAWC~J1857+027}                 &   507.4 & 284.33 &   2.80 &  36.09 &  -0.03 &  0.06  & 3HWC~J1857+027, \textbf{HESS J1857+026}\\
 \color{orange} \textbf{HAWC~J1859+057$^{\dag}$} &   33.0  & 284.85 &   5.72 &  38.93 &   0.84 &  0.13  & 3HWC J1857+051$^{\dag}$ \\
 \color{orange} \textbf{HAWC~J1907+085}  &   48.2  & 286.79 &   8.57 &  42.35 &   0.44 &  0.10  & 3HWC J1907+085\\
 \textbf{HAWC~J1908+063}                 &  1068.5 & 287.05 &   6.39 &  40.53 &  -0.80 &  0.06  & 3HWC~J1908+063, \textbf{HESS J1908+063}, \\
                                         &         &        &        &        &        &        & MGRO J1908+06 \\
 \textbf{HAWC~J1912+103}                 &   130.8 & 288.02 &  10.39 &  44.52 &   0.20 &  0.07  & 3HWC~J1912+103, \textbf{HESS J1912+101}\\
 \color{orange} \textbf{HAWC~J1914+118}  &   64.2  & 288.68 &  11.87 &  46.13 &   0.32 &  0.09  & 3HWC J1914+118\\
 \color{orange} \textbf{HAWC~J1920+147}  &   34.4  & 290.17 &  14.79 &  49.39 &   0.39 &  0.10   & 3HWC~J1920+147$^{\dag}$, W 51 \\
 \textbf{HAWC~J1922+140}                 &   138.1 & 290.70 &  14.09 &  49.01 &  -0.38 &  0.06 & 3HWC~J1922+140, \textbf{HESS J1923+140}, W 51  \\
 \color{orange} \textbf{HAWC~J1923+169$^{\dag}$} &   28.2  & 290.79 &  16.96 &  51.58 &   0.89 &  0.11  & 3HWC~J1923+169$^{\dag}$ \\
 \color{orange} \textbf{HAWC~J1928+178}  &   172.5 & 292.10 &  17.82 &  52.93 &   0.20 &  0.09  & 3HWC J1928+178\\
 \textbf{HAWC~J1930+188}                 &    86.1 & 292.54 &  18.84 &  54.03 &   0.32 &  0.09 & 3HWC~J1930+188, \textbf{HESS J1930+188}, \\
                                         &         &        &        &        &        &        & SNR G054.1+00.3 \\
  \hline
 \end{tabular}
\end{table*}

\newpage

\begin{table*}[h!]
  \centering
 \caption{HAWC source list with the HAWC and H.E.S.S. integrated flux above 1 TeV corresponding to this analysis.
 The flux values are reported at the location of the HAWC sources from Table~\ref{tab:source1}. The H.E.S.S. flux from this analysis is also reported at the location of the HGPS source when there is an association. The positions of the HGPS sources are given in galactic coordinates into parentheses. The flux uncertainty reported here is statistical only. The systematic uncertainties on the flux are estimated to be 30\% for H.E.S.S.~\citep{HGPS} and 30\% for HAWC~\citep{HAWC_crab_100TeV}. Note that the HAWC flux reported here is slightly different from the output from the HAWC source search form~\citet{3HWC_catalog} on the data set considered here because it assumes a simple power law spectrum with a constant spectral index of $-2.5$ for all sources, while in the catalogue search it is a free parameter from the fit.   
 Moreover, the flux reported for H.E.S.S. is expected to differ from the one reported in the HGPS due to the hypotheses of 0.4\textdegree \ on the gaussian size, and the use of an spectral index of $-2.5$ instead of $-2.3$ and the different background estimation method} 
 \label{tab:source2}.
 \begin{tabular}{c c c c c}
  \hline
  \hline
  Name           & F$_{\mbox{HAWC >1TeV}}$               & F$_{\mbox{HESS >1TeV}}$      & HGPS counterpart ($\ell$, $b$)  & F$_{\mbox{HESS >1TeV}}$ at HGPS location   \\
                 & (10$^{-12}$ cm$^{-2}$s$^{-1}$)          & (10$^{-12}$ cm$^{-2}$s$^{-1}$) &                   & (10$^{-12}$ cm$^{-2}$s$^{-1}$))                \\
 \hline
 \textbf{HAWC J1809-190}  & $5.98 \pm 0.47$     & $4.15 \pm 0.17$  & \textbf{HESS J1809-193} (11.11, -0.02) & $5.37 \pm 0.45$\\
 \textbf{HAWC J1813-125}  & $1.29 \pm 0.22$ & $1.11 \pm 0.25$ & \textbf{HESS J1813-126} (17.31, 2.49) &  $1.04 \pm  0.21$ \\
 \textbf{HAWC J1813-174}  & $6.49 \pm 0.41$    & $5.74 \pm 0.33$ & \textbf{HESS J1813-178} (12.82, -0.025) & $2.12 \pm 0.40$ \\
 \textbf{HAWC J1819-151}  & $2.13 \pm 0.28$     & $1.16 \pm 0.17$ & \textbf{HESS J1818-154} (15.41, 0.16) & $0.23 \pm 0.05$ \\
 \textbf{HAWC J1825-134}  & $11.4 \pm 0.34$     & $9.78 \pm 0.20$ & \textbf{HESS J1825-137} (17.52, -0.62) & $19.15 \pm 1.85$ \\
 \textbf{HAWC J1831-096}  & $2.29 \pm 0.20$     & $2.28 \pm 0.18$ & - & - \\
 \textbf{HAWC J1837-066}  & $5.65 \pm 0.21$     & $7.32 \pm 0.25$ & \textbf{HESS J1837-069} (25.15, -0.087) & $11.55\pm 0.49$ \\
 \textbf{HAWC J1843-034}  & $3.64 \pm 0.16$     & $4.27 \pm 0.24$ & \textbf{HESS J1843-033} (28.90, 0.075) & $3.04\pm 0.20$ \\
 \textbf{HAWC J1847-017}  & $1.82 \pm 0.14$     & $2.66 \pm 0.19$ & \textbf{HESS J1848-018} (30.92, -0.20) & $1.11 \pm 0.15$  \\
 \textbf{HAWC J1849+001}  & $2.11 \pm 0.13$     & $2.03 \pm 0.17$ & \textbf{HESS J1849-000} (32.61, 0.53) & $0.57 \pm 0.07$ \\
  \color{orange} \textbf{HAWC J1852+013}  & $0.97 \pm 0.11$    & $1.22 \pm 0.16$ & -  & - \\
 \textbf{HAWC J1857+027}  & $2.54 \pm 0.13$     & $4.75 \pm 0.26$  & \textbf{HESS J1857+026} (36.06, -0.06) & $4.0 \pm 0.29$ \\
  \color{orange} \textbf{HAWC J1859+057$^{\dag}$}  & $0.48 \pm 0.091$   & $0.39 \pm 0.31$ & - & - \\
  \color{orange} \textbf{HAWC J1907+085}  & $0.56 \pm 0.089$    & $1.03 \pm 0.16$ & - & - \\
 \textbf{HAWC J1908+063}  & $3.52 \pm 0.13$     & $3.71 \pm 0.22$ & \textbf{HESS J1908+063} (40.55, -0.84) & $8.35 \pm 0.57$ \\
 \textbf{HAWC J1912+103}  & $0.98 \pm 0.096$    & $2.01 \pm 0.13$ & \textbf{HESS J1912+101} (44.46 -0.13) & $2.49 \pm 0.34$ \\
  \color{orange} \textbf{HAWC J1914+118}  & $0.61 \pm 0.085$    & $1.09 \pm 0.20$ & - & - \\
  \color{orange} \textbf{HAWC J1920+147}  & $0.43 \pm 0.080$   & $0.54 \pm 0.22$ & - & - \\
 \textbf{HAWC J1922+140}  & $0.92 \pm 0.088$    & $1.38 \pm 0.22$ & \textbf{HESS J1923+141} (49.08, -0.40) & $0.69 \pm 0.11$ \\
  \color{orange} \textbf{HAWC J1923+169$^{\dag}$}  & $0.39 \pm 0.079$   & $0.51 \pm 0.31$ & - & - \\
  \color{orange} \textbf{HAWC J1928+178}  & $1.01 \pm 0.09$     & $1.49 \pm 0.24$ & - & - \\
 \textbf{HAWC J1930+188}  & $0.74 \pm 0.087$    & $1.77 \pm 0.23$ & \textbf{HESS J1930+188} (54.06, 0.27) & $0.32 \pm 0.068$ \\

  \hline
 \end{tabular}
\end{table*}

\newpage 

\section{Systematic uncertainties}
\label{Systematics}

The difference between HAWC and H.E.S.S. flux profiles may be explained, at least partially, by evaluating the systematic uncertainties. First, each instrument has its own absolute systematic uncertainties on the flux: for H.E.S.S., they are mainly due to atmospheric effects (weather condition and seasonal effect), but also to the modeling of the instrument, and have been evaluated to $\sim$30\%~\citep{HGPS}. HAWC is dominated by the uncertainties on the modeling of the instrument, which results in $\sim$30\% uncertainty on the flux \citep{HAWC_crab_100TeV}. Secondly, the variation in flux resulting from a variation of each of the parameters listed bellow, used to produce the maps, is evaluated. For most of them, the median flux variation over the part of the Galactic plane used for the comparison ($10\mbox{\textdegree} < \ell < 60\mbox{\textdegree}$) and local effects may have to be taken into account for specific studies. When a parameter highly depends on the longitude, the flux variation is derived for all positions along the Galactic plane. All systematic uncertainties are then added quadratically and represented as the dotted envelope in Figure~\ref{long_profile}.
\begin{itemize}
    \item The spectral index: An index of $-2.5$ is used both for HAWC and H.E.S.S. for the entire part of the Galactic plane considered for the comparison. However, H.E.S.S. used $-2.3$ when producing the HGPS, as being the mean index of teraelectronvolt Galactic sources. Another H.E.S.S. map was produced (not presented here) using the new analysis presented in this paper and an index of $-2.3$. Using an index of $-2.3$ instead of $-2.5$ leads to a median flux variation of $\sim$1.7\% over the part of the Galactic plane considered here.
    \item The analysis energy threshold: While it can be set quite precisely to be 1~TeV for H.E.S.S. data, with an uncertainty of $\sim$10\%~\citep{ImPACT}, the uncertainty on the HAWC energy threshold is much larger due to the use of the analysis bins. This can cause HAWC to be dominated by events of a few teraelectronvolts instead of 1~TeV, depending on the zenith angle. A H.E.S.S. map was produced using the new analysis presented in this paper and an energy threshold of 2~TeV instead of 1~TeV. Using 2~TeV would lead to a median variation of $\sim$14\%. 
    \item The background method: A H.E.S.S. map was produced (not presented here) using the new analysis presented in this paper but using the ring background method. Using the field-of-view background method instead of the ring background method leads to a systematic increase of the flux by a median value of $1.9\times 10^{-13}$~cm$^{-2}$~s$^{-1}$.
    \item The imperfect knowledge of the H.E.S.S. background, due to the local nature of the background estimation and the sensitivity to instrumental conditions: Assuming that the background is known with a 5\% precision, the corresponding error on the flux at each point in longitude is evaluated to range between $1.5\times 10^{-13}$~cm$^{-2}$~s$^{-1}$ at $\ell~=~10^{\circ}$ and $5\times 10^{-13}$~cm$^{-2}$~s$^{-1}$ at $\ell~=~60^{\circ}$.
    \item H.E.S.S. analysis chain: A H.E.S.S. map was produced (not presented here) using the new analysis presented in this paper with the other H.E.S.S. analysis chain to cross check the results. The median variation in flux seen by the two different H.E.S.S. analysis chains is $\sim$25\%. % of the flux.
    \item Using the HAWC flux map: In the study presented in this paper, the flux are obtained by reading the flux map, making the assumption that the spectral index is $-2.5$ for all sources. The flux values can be compared with the one obtained from the HAWC catalog search, where both the index and the flux normalization are fitted. The median difference in flux between both is $\sim$48\% of the flux.
    \item The source size hypotheses: A constant size of 0.4\textdegree \ was used in this analysis to produce the H.E.S.S. TS map, whereas the PSF of HAWC varies with zenith angle and as a function of the analysis bin. The average HAWC PSF over the analysis bins 4 to 9 is actually closer to 0.3$^{\circ}$. Another H.E.S.S. map using the new analysis presented in this paper was produced (not shown here) using a Gaussian of 0.3\textdegree \ instead of 0.4\textdegree. Decreasing the size of the Gaussian to 0.3$^{\circ}$ leads to a flux $39\%$ lower.
    \item The systematic bias on the source position reconstruction (visible in Figure~11 of~\citet{3HWC_catalog}) has also been taken into account in the systematic uncertainties: Similarly, a relation has been derived to describe the difference in the reconstructed location of the sources in common to the HAWC data set used this analysis and the HGPS (Table~\ref{tab:source2}). Multiple profiles have then been derived and averaged according to this relation and its errors, given a mean systematic uncertainty that depends on the longitude. The systematic error is then estimated to be $9.3\times 10^{-14}$~cm$^{-2}$~s$^{-1}$ (median), which is negligible compare to the other systematic uncertainties. 
\end{itemize}

An in-depth study with joint analysis of common sources would be needed to quantify these effects more precisely; this is beyond the scope of the first approach presented here.

\newpage

\section{Latitude profiles in 10\textdegree \ longitude bands}
\label{more_latitude_profile}

\begin{figure*}[ht!]
   \centering
   \includegraphics[width=1\linewidth]{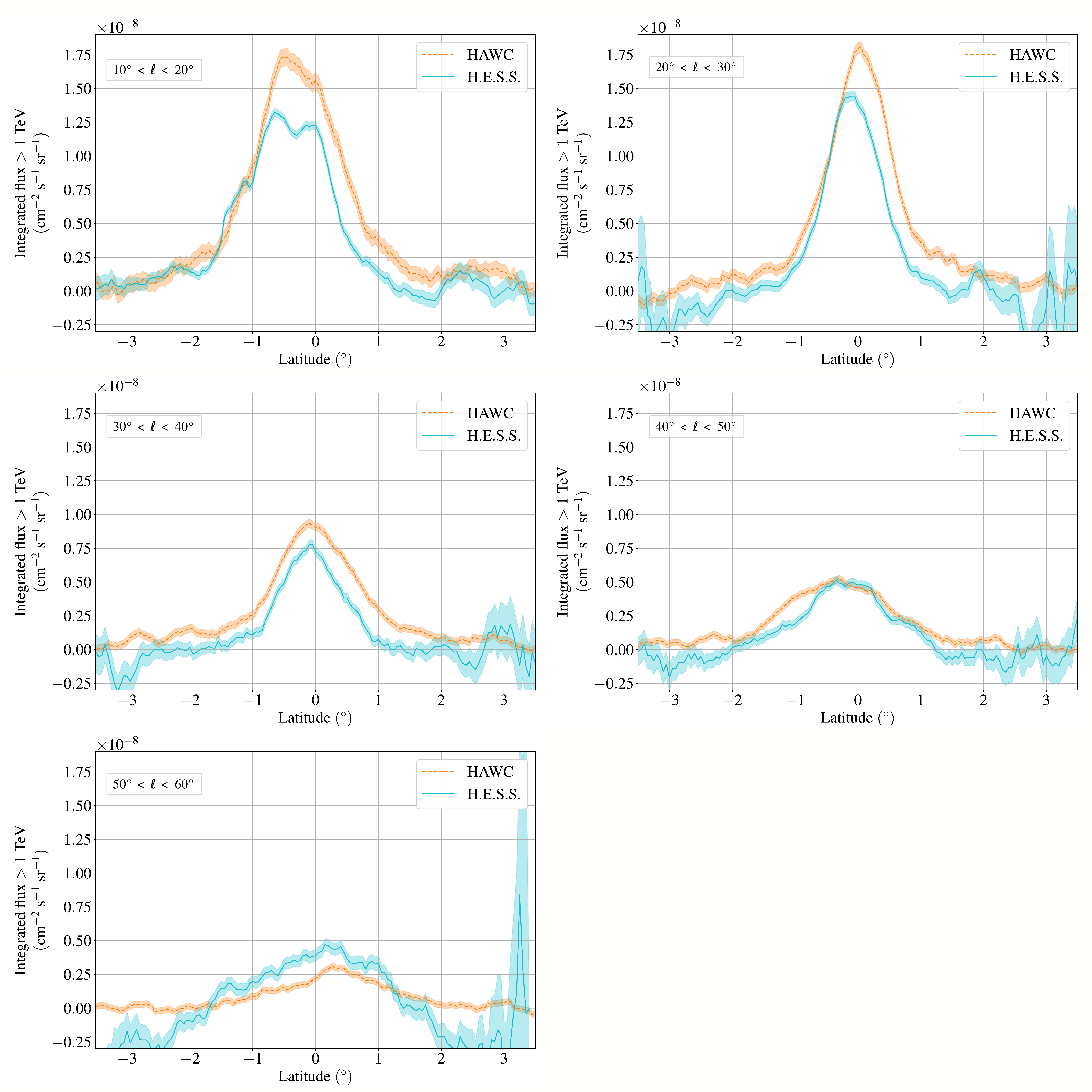}
    \caption{Profiles of the integrated flux above 1~TeV for H.E.S.S. and HAWC along the latitude axis in bands of 10\textdegree \ in longitude. The shaded bands represent the 1$\sigma$ uncertainties. The systematic uncertainties are not shown here.
    }
   \label{fig:lat_profile_10deg}
   \end{figure*}

The latitude profiles of the integrated flux above 1~TeV for H.E.S.S. and HAWC, as described in section~\ref{profiles}, are plotted in Figure~\ref{fig:lat_profile_10deg} in bands of 10\textdegree \ in longitude as indicated on the plots. HAWC detects more flux than H.E.S.S. in all bands with $\ell < 40$\textdegree. In the band with $40^{\circ} < \ell < 50^{\circ}$ HAWC and H.E.S.S. detect a similar level of flux, and for $50^{\circ} < \ell < 60^{\circ}$, H.E.S.S. detects slightly more flux. However, when taking into account the exposure time of each instrument over each latitude range, in the average over $10^{\circ} < \ell < 60^{\circ}$, H.E.S.S. detects more flux as shown in the left plot of Figure~\ref{fig:lat_profile}. Note that the systematic uncertainties are not shown here but can reach 50\% of the flux. 

\section{Applicability of Wilk's theorem on HAWC and H.E.S.S. data}
\label{mapsimu}

\begin{figure*}[ht!]
   \centering
   \includegraphics[width=0.78\linewidth]{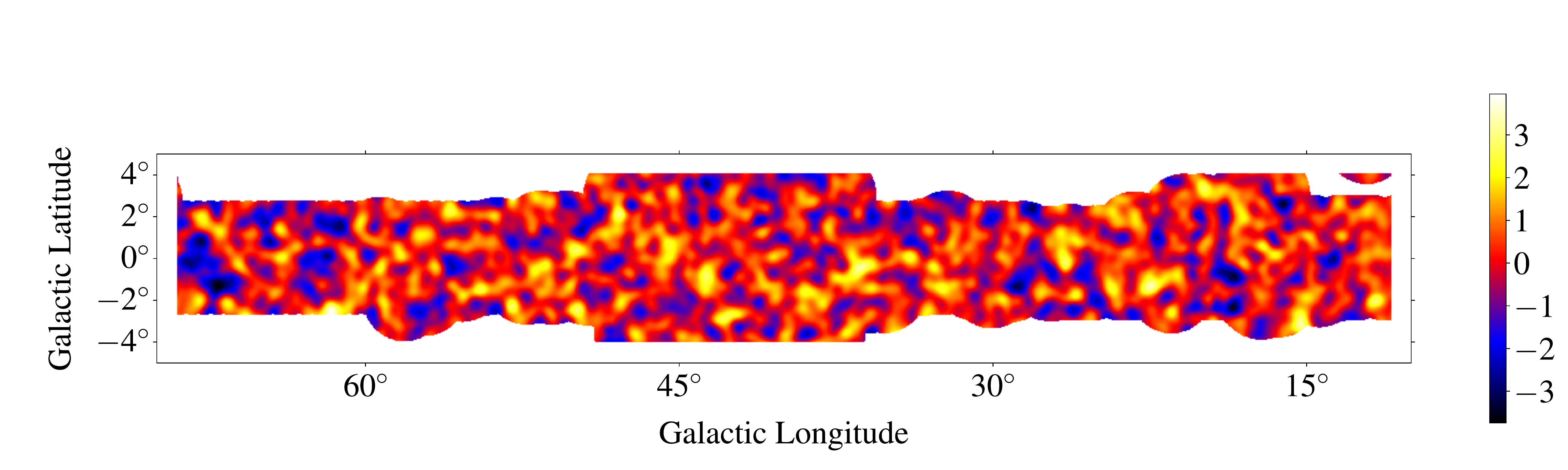}
   \includegraphics[width=0.2\linewidth]{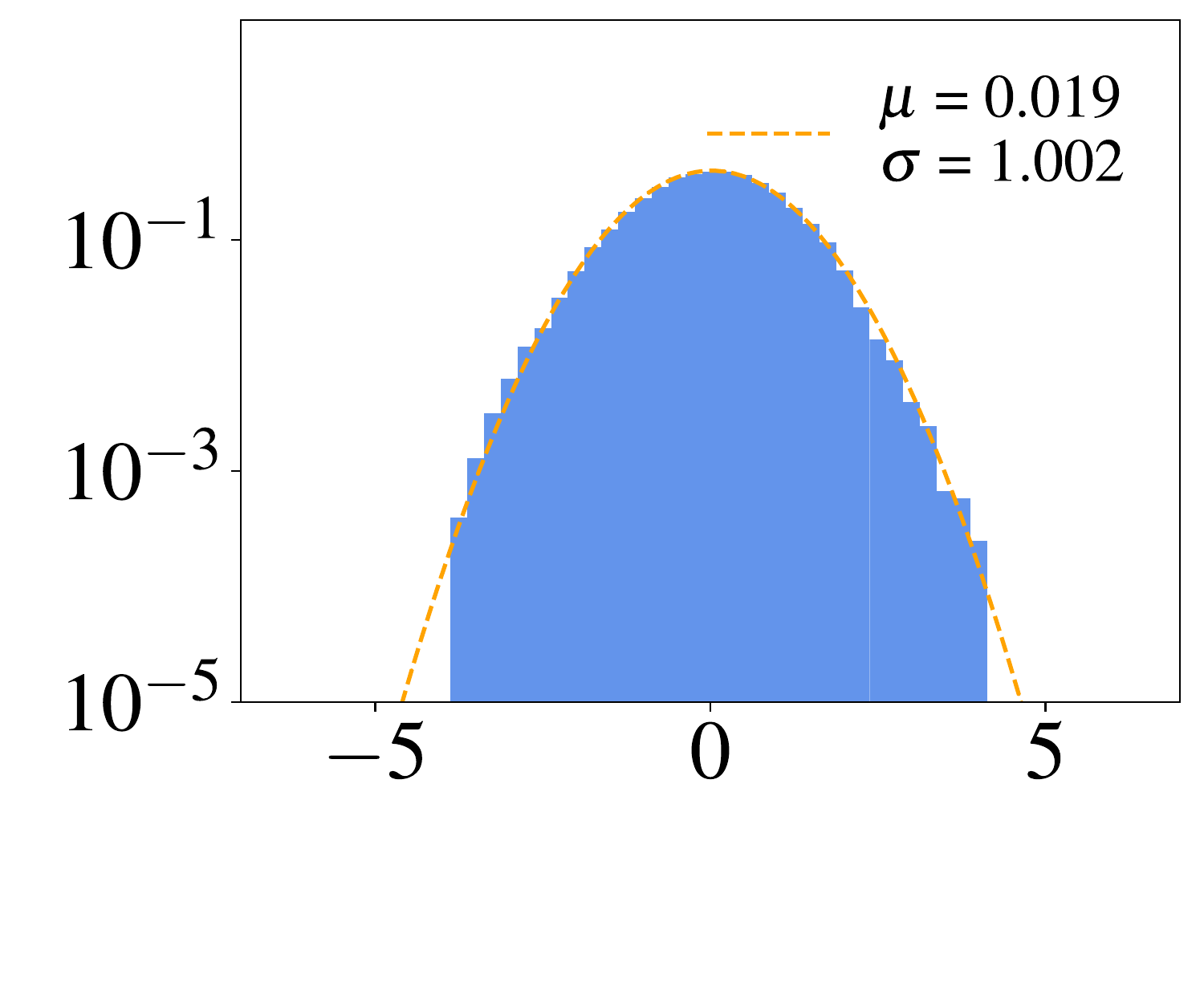}
    \includegraphics[width=0.78\linewidth]{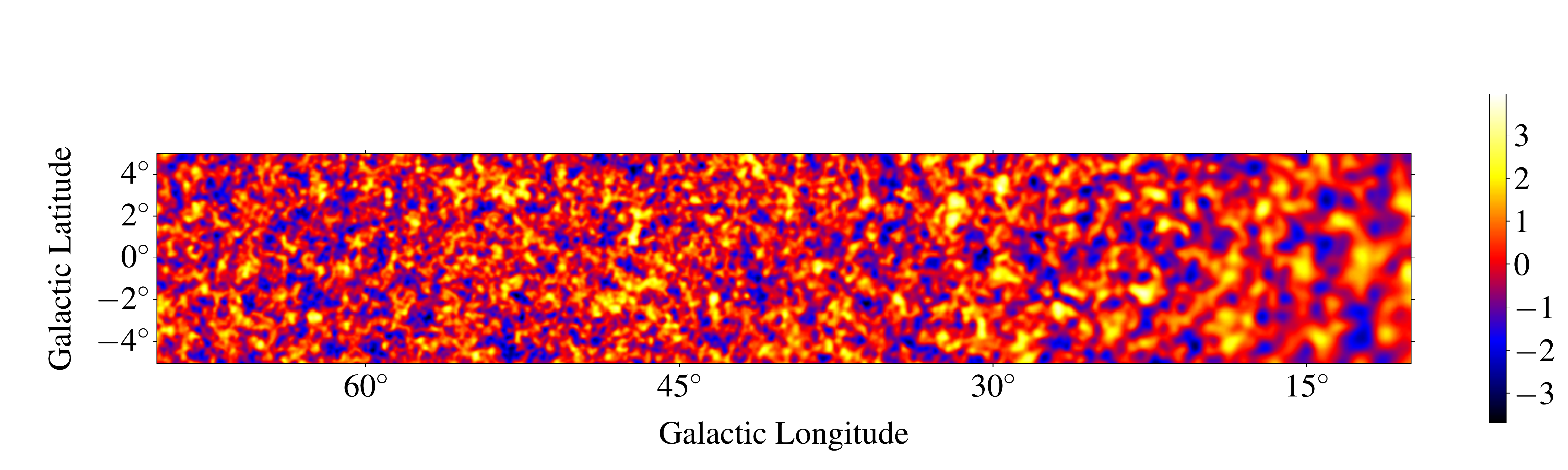}
   \includegraphics[width=0.2\linewidth]{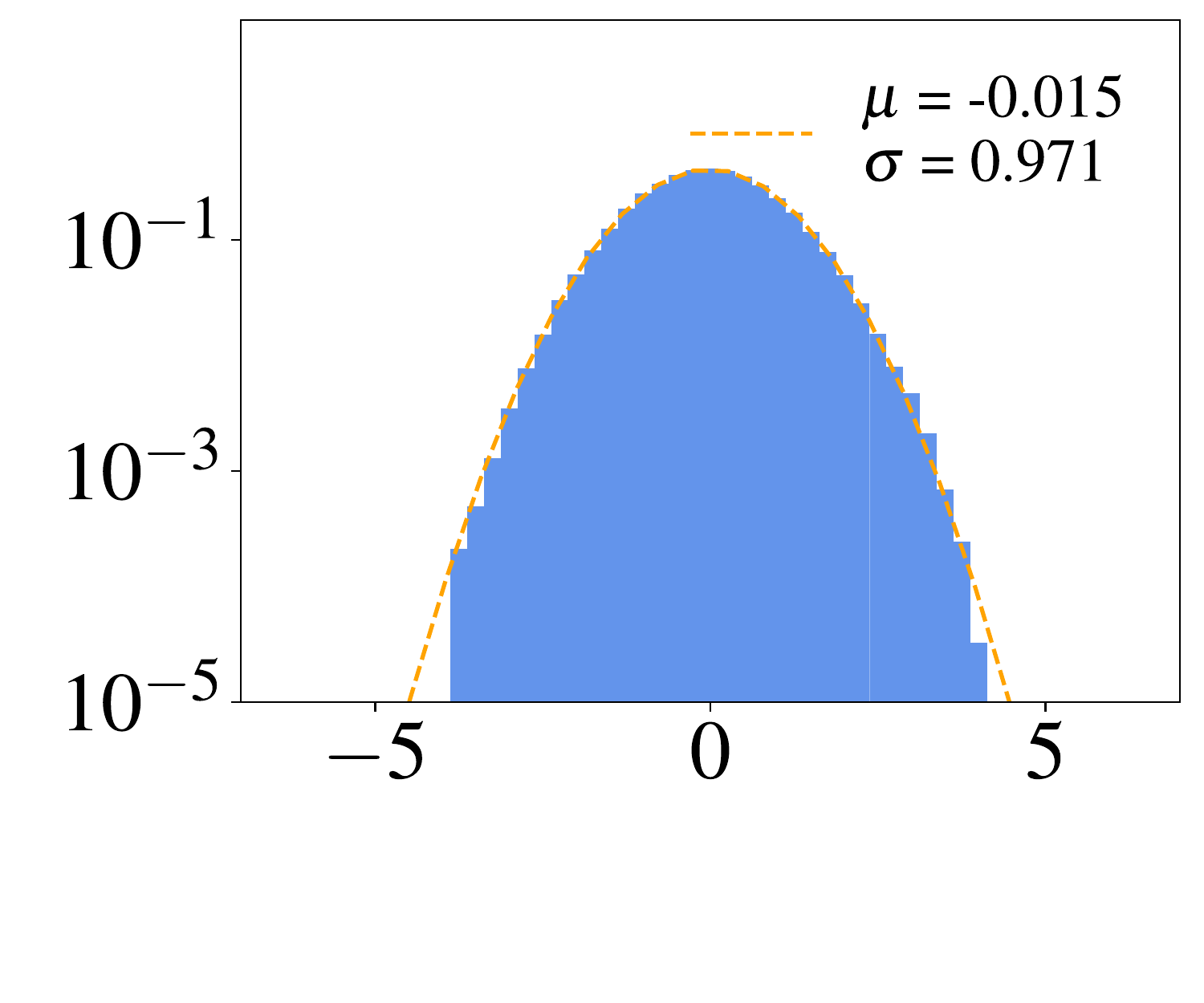}
    \caption{H.E.S.S. (top) and HAWC (bottom) synthetic significance map with their distribution.
    }
   \label{fake_map}
   \end{figure*}

To validate the use of Wilk's theorem and check the significance distribution, synthetic maps have been created for HAWC and H.E.S.S. by generating random Poisson realizations assuming background only. The significance maps are then produced in the same way as the maps presented in the paper, as described in section~\ref{analysis}.
Figure~\ref{fake_map} shows an example for one synthetic map for H.E.S.S. with the new analysis and one for HAWC using the standard analysis. 
The distribution of significance in the resulting null hypothesis maps is found to be consistent with a Gaussian centered on 0 with a standard deviation of 1, as expected. 

%% For this sample we use BibTeX plus aasjournals.bst to generate the
%% the bibliography. The sample63.bib file was populated from ADS. To
%% get the citations to show in the compiled file do the following:
%%
%% pdflatex sample63.tex
%% bibtext sample63
%% pdflatex sample63.tex
%% pdflatex sample63.tex

\newpage 

\bibliography{biblio}{}

\begin{thebibliography}{}
\expandafter\ifx\csname natexlab\endcsname\relax\def\natexlab#1{#1}\fi
\providecommand{\url}[1]{\href{#1}{#1}}
\providecommand{\dodoi}[1]{doi:~\href{http://doi.org/#1}{\nolinkurl{#1}}}
\providecommand{\doeprint}[1]{\href{http://ascl.net/#1}{\nolinkurl{http://ascl.net/#1}}}
\providecommand{\doarXiv}[1]{\href{https://arxiv.org/abs/#1}{\nolinkurl{https://arxiv.org/abs/#1}}}

\bibitem[{{Abdo} {et~al.}(2007){Abdo}, {Allen}, {Berley}, {Casanova}, {Chen},
  {Coyne}, {Dingus}, {Ellsworth}, {Fleysher}, {Fleysher}, {Gonzalez},
  {Goodman}, {Hays}, {Hoffman}, {Hopper}, {H{\"u}ntemeyer}, {Kolterman},
  {Lansdell}, {Linnemann}, {McEnery}, {Mincer}, {Nemethy}, {Noyes}, {Ryan},
  {Saz Parkinson}, {Shoup}, {Sinnis}, {Smith}, {Sullivan}, {Vasileiou},
  {Walker}, {Williams}, {Xu}, \& {Yodh}}]{MGROJ1908+06}
{Abdo}, A.~A., {Allen}, B., {Berley}, D., {et~al.} 2007, \apjl, 664, L91,
  \dodoi{10.1086/520717}

\bibitem[{{Abeysekara} {et~al.}(2017{\natexlab{a}}){Abeysekara}, {Albert},
  {Alfaro}, {Alvarez}, {{\'A}lvarez}, {Arceo}, {Arteaga-Vel{\'a}zquez}, {Ayala
  Solares}, {Barber}, {Bautista-Elivar}, {Becerril}, {Belmont-Moreno},
  {BenZvi}, {Berley}, {Braun}, {Brisbois}, {Caballero-Mora}, {Capistr{\'a}n},
  {Carrami{\~n}ana}, {Casanova}, {Castillo}, {Cotti}, {Cotzomi}, {Couti{\~n}o
  de Le{\'o}n}, {de la Fuente}, {De Le{\'o}n}, {DeYoung}, {Dingus},
  {DuVernois}, {D{\'{\i}}az-V{\'e}lez}, {Ellsworth}, {Fiorino}, {Fraija},
  {Garc{\'{\i}}a-Gonz{\'a}lez}, {Gerhardt}, {Gonz{\'a}lez Mun{\"o}z},
  {Gonz{\'a}lez}, {Goodman}, {Hampel-Arias}, {Harding}, {Hernandez},
  {Hernandez-Almada}, {Hinton}, {Hui}, {H{\"u}ntemeyer}, {Iriarte},
  {Jardin-Blicq}, {Joshi}, {Kaufmann}, {Kieda}, {Lara}, {Lauer}, {Lee},
  {Lennarz}, {Le{\'o}n Vargas}, {Linnemann}, {Longinotti}, {Raya},
  {Luna-Garc{\'{\i}}a}, {L{\'o}pez-Coto}, {Malone}, {Marinelli}, {Martinez},
  {Martinez-Castellanos}, {Mart{\'{\i}}nez-Castro}, {Mart{\'{\i}}nez-Huerta},
  {Matthews}, {Miranda-Romagnoli}, {Moreno}, {Mostaf{\'a}}, {Nellen},
  {Newbold}, {Nisa}, {Noriega-Papaqui}, {Pelayo}, {Pretz},
  {P{\'e}rez-P{\'e}rez}, {Ren}, {Rho}, {Rivi{\`e}re}, {Rosa-Gonz{\'a}lez},
  {Rosenberg}, {Ruiz-Velasco}, {Salazar}, {Salesa Greus}, {Sandoval},
  {Schneider}, {Schoorlemmer}, {Sinnis}, {Smith}, {Springer}, {Surajbali},
  {Taboada}, {Tibolla}, {Tollefson}, {Torres}, {Ukwatta}, {Villase{\~n}or},
  {Weisgarber}, {Westerhoff}, {Wisher}, {Wood}, {Yapici}, {Yodh}, {Younk},
  {Zepeda}, \& {Zhou}}]{HAWC_crab}
{Abeysekara}, A.~U., {Albert}, A., {Alfaro}, R., {et~al.} 2017{\natexlab{a}},
  The Astrophysical Journal, 843, 39, \dodoi{10.3847/1538-4357/aa7555}

\bibitem[{{Abeysekara} {et~al.}(2017{\natexlab{b}}){Abeysekara}, {Albert},
  {Alfaro}, {Alvarez}, {{\'A}lvarez}, {Arceo}, {Arteaga-Vel{\'a}zquez}, {Ayala
  Solares}, {Barber}, {Baughman}, {Bautista-Elivar}, {Becerra Gonzalez},
  {Becerril}, {Belmont-Moreno}, {BenZvi}, {Berley}, {Bernal}, {Braun},
  {Brisbois}, {Caballero-Mora}, {Capistr{\'a}n}, {Carrami{\~n}ana}, {Casanova},
  {Castillo}, {Cotti}, {Cotzomi}, {Couti{\~n}o de Le{\'o}n}, {de la Fuente},
  {De Le{\'o}n}, {Diaz Hernandez}, {Dingus}, {DuVernois},
  {D{\'{\i}}az-V{\'e}lez}, {Ellsworth}, {Engel}, {Fiorino}, {Fraija},
  {Garc{\'{\i}}a-Gonz{\'a}lez}, {Garfias}, {Gerhardt}, {Gonz{\'a}lez
  Mu{\~n}oz}, {Gonz{\'a}lez}, {Goodman}, {Hampel-Arias}, {Harding},
  {Hernandez}, {Hernandez-Almada}, {Hinton}, {Hui}, {H{\"u}ntemeyer},
  {Iriarte}, {Jardin-Blicq}, {Joshi}, {Kaufmann}, {Kieda}, {Lara}, {Lauer},
  {Lee}, {Lennarz}, {Le{\'o}n Vargas}, {Linnemann}, {Longinotti}, {Raya},
  {Luna-Garc{\'{\i}}a}, {L{\'o}pez-Coto}, {Malone}, {Marinelli}, {Martinez},
  {Martinez-Castellanos}, {Mart{\'{\i}}nez-Castro}, {Mart{\'{\i}}nez-Huerta},
  {Matthews}, {Miranda-Romagnoli}, {Moreno}, {Mostaf{\'a}}, {Nellen},
  {Newbold}, {Nisa}, {Noriega-Papaqui}, {Pelayo}, {Pretz},
  {P{\'e}rez-P{\'e}rez}, {Ren}, {Rho}, {Rivi{\`e}re}, {Rosa-Gonz{\'a}lez},
  {Rosenberg}, {Ruiz-Velasco}, {Salazar}, {Salesa Greus}, {Sandoval},
  {Schneider}, {Schoorlemmer}, {Sinnis}, {Smith}, {Springer}, {Surajbali},
  {Taboada}, {Tibolla}, {Tollefson}, {Torres}, {Ukwatta}, {Vianello},
  {Villase{\~n}or}, {Weisgarber}, {Westerhoff}, {Wisher}, {Wood}, {Yapici},
  {Younk}, {Zepeda}, \& {Zhou}}]{HAWC_catalog}
---. 2017{\natexlab{b}}, The Astrophysical Journal, 843, 40,
  \dodoi{10.3847/1538-4357/aa7556}

\bibitem[{{Abeysekara} {et~al.}(2018){Abeysekara}, {Archer}, {Benbow}, {Bird},
  {Brose}, {Buchovecky}, {Buckley}, {Bugaev}, {Chromey}, {Connolly}, {Cui},
  {Daniel}, {Falcone}, {Feng}, {Finley}, {Fortson}, {Furniss}, {H{\"u}tten},
  {Hanna}, {Hervet}, {Holder}, {Hughes}, {Humensky}, {Johnson}, {Kaaret},
  {Kar}, {Kertzman}, {Kieda}, {Krause}, {Krennrich}, {Kumar}, {Lang}, {Lin},
  {McArthur}, {Moriarty}, {Mukherjee}, {O'Brien}, {Ong}, {Otte}, {Park},
  {Petrashyk}, {Pohl}, {Pueschel}, {Quinn}, {Ragan}, {Reynolds}, {Richards},
  {Roache}, {Rulten}, {Sadeh}, {Santander}, {Sembroski}, {Shahinyan}, {Sushch},
  {Tyler}, {Wakely}, {Weinstein}, {Wells}, {Wilcox}, {Wilhelm}, {Williams},
  {Williamson}, {Zitzer}, {VERITAS Collaboration}, {Abdollahi}, {Ajello},
  {Baldini}, {Barbiellini}, {Bastieri}, {Bellazzini}, {Berenji}, {Bissaldi},
  {Bland ford}, {Bonino}, {Bottacini}, {Brandt}, {Bruel}, {Buehler}, {Cameron},
  {Caputo}, {Caraveo}, {Castro}, {Cavazzuti}, {Charles}, {Chiaro}, {Ciprini},
  {Cohen-Tanugi}, {Costantin}, {Cutini}, {D'Ammand o}, {de Palma}, {Di Lalla},
  {Di Mauro}, {Di Venere}, {Dom{\'\i}nguez}, {Favuzzi}, {Fegan}, {Franckowiak},
  {Fukazawa}, {Funk}, {Fusco}, {Gargano}, {Gasparrini}, {Giglietto},
  {Giordano}, {Giroletti}, {Green}, {Grenier}, {Guillemot}, {Guiriec}, {Hays},
  {Hewitt}, {Horan}, {J{\'o}hannesson}, {Kensei}, {Kuss}, {Larsson},
  {Latronico}, {Lemoine-Goumard}, {Li}, {Longo}, {Loparco}, {Lovellette},
  {Lubrano}, {Magill}, {Maldera}, {Mazziotta}, {McEnery}, {Michelson},
  {Mitthumsiri}, {Mizuno}, {Monzani}, {Morselli}, {Moskalenko}, {Negro},
  {Nuss}, {Ojha}, {Omodei}, {Orienti}, {Orlando}, {Palatiello}, {Paliya},
  {Paneque}, {Perkins}, {Persic}, {Pesce-Rollins}, {Petrosian}, {Piron},
  {Porter}, {Principe}, {Rain{\`o}}, {Rando}, {Rani}, {Razzano}, {Razzaque},
  {Reimer}, {Reimer}, {Reposeur}, {Sgr{\`o}}, {Siskind}, {Spandre}, {Spinelli},
  {Suson}, {Tajima}, {Thayer}, {Thompson}, {Torres}, {Tosti}, {Troja},
  {Valverde}, {Vianello}, {Vogel}, {Wood}, {Yassine}, {Fermi-LAT
  Collaboration}, {Alfaro}, {{\'A}lvarez}, {{\'A}lvarez}, {Arceo},
  {Arteaga-Vel{\'a}zquez}, {Avila Rojas}, {Ayala Solares}, {Becerril},
  {Belmont-Moreno}, {BenZvi}, {Bernal}, {Braun}, {Brisbois}, {Caballero-Mora},
  {Capistr{\'a}n}, {Carrami{\~n}ana}, {Casanova}, {Castillo}, {Cotti},
  {Cotzomi}, {Couti{\~n}o de Le{\'o}n}, {De Le{\'o}n}, {De la Fuente},
  {Dichiara}, {Dingus}, {DuVernois}, {D{\'\i}az-V{\'e}lez}, {Engel},
  {Enriquez-Rivera}, {Fiorino}, {Fleischhack}, {Fraija},
  {Garc{\'\i}a-Gonz{\'a}lez}, {Garfias}, {Gonz{\'a}lez Mu{\~n}oz},
  {Gonz{\'a}lez}, {Goodman}, {Hampel-Arias}, {Harding}, {Hernand ez},
  {Hernandez-Almada}, {Hona}, {Hueyotl-Zahuantitla}, {Hui}, {H{\"u}ntemeyer},
  {Iriarte}, {Jardin-Blicq}, {Joshi}, {Kaufmann}, {Lara}, {Lauer}, {Lee},
  {Lennarz}, {Le{\'o}n Vargas}, {Linnemann}, {Longinotti}, {Luis-Raya},
  {Luna-Garc{\'\i}a}, {L{\'o}pez-Coto}, {Malone}, {Marinelli}, {Martinez},
  {Martinez-Castellanos}, {Mart{\'\i}nez-Castro}, {Mart{\'\i}nez-Huerta},
  {Matthews}, {Miranda-Romagnoli}, {Moreno}, {Mostaf{\'a}}, {Nayerhoda},
  {Nellen}, {Newbold}, {Nisa}, {Noriega-Papaqui}, {Pelayo}, {Pretz},
  {P{\'e}rez-P{\'e}rez}, {Ren}, {Rho}, {Rivi{\`e}re}, {Rosa-Gonz{\'a}lez},
  {Rosenberg}, {Ruiz-Velasco}, {Salazar}, {Salesa Greus}, {Sandoval},
  {Schneider}, {Seglar Arroyo}, {Sinnis}, {Smith}, {Springer}, {Surajbali},
  {Taboada}, {Tibolla}, {Tollefson}, {Torres}, {Ukwatta}, {Villase{\~n}or},
  {Weisgarber}, {Westerhoff}, {Wisher}, {Wood}, {Yapici}, {Yodh}, {Zepeda},
  {Zhou}, \& {HAWC Collaboration}}]{HAWC_Fermi_VERITAS}
{Abeysekara}, A.~U., {Archer}, A., {Benbow}, W., {et~al.} 2018, \apj, 866, 24,
  \dodoi{10.3847/1538-4357/aade4e}

\bibitem[{{Abeysekara} {et~al.}(2019){Abeysekara}, {Albert}, {Alfaro},
  {Alvarez}, {{\'A}lvarez}, {Camacho}, {Arceo}, {Arteaga-Vel{\'a}zquez},
  {Arunbabu}, {Avila Rojas}, {Ayala Solares}, {Baghmanyan}, {Belmont-Moreno},
  {BenZvi}, {Brisbois}, {Caballero-Mora}, {Capistr{\'a}n}, {Carrami{\~n}ana},
  {Casanova}, {Cotti}, {Cotzomi}, {Couti{\~n}o de Le{\'o}n}, {De la Fuente},
  {de Le{\'o}n}, {Dichiara}, {Dingus}, {DuVernois}, {D{\'\i}az-V{\'e}lez},
  {Ellsworth}, {Engel}, {Espinoza}, {Fick}, {Fleischhack}, {Fraija},
  {Galv{\'a}n-G{\'a}mez}, {Garc{\'\i}a-Gonz{\'a}lez}, {Garfias},
  {Gonz{\'a}lez}, {Goodman}, {Harding}, {Hernandez}, {Hinton}, {Hona},
  {Hueyotl-Zahuantitla}, {Hui}, {H{\"u}ntemeyer}, {Iriarte}, {Jardin-Blicq},
  {Joshi}, {Kaufmann}, {Kieda}, {Lara}, {Lee}, {Le{\'o}n Vargas}, {Linnemann},
  {Longinotti}, {Luis-Raya}, {Lundeen}, {Malone}, {Marinelli}, {Martinez},
  {Martinez-Castellanos}, {Mart{\'\i}nez-Castro}, {Mart{\'\i}nez-Huerta},
  {Matthews}, {Miranda-Romagnoli}, {Morales-Soto}, {Moreno}, {Mostaf{\'a}},
  {Nayerhoda}, {Nellen}, {Newbold}, {Nisa}, {Noriega-Papaqui}, {Peisker},
  {P{\'e}rez-P{\'e}rez}, {Pretz}, {Ren}, {Rho}, {Rivi{\`e}re},
  {Rosa-Gonz{\'a}lez}, {Rosenberg}, {Ruiz-Velasco}, {Salazar}, {Salesa Greus},
  {Sandoval}, {Schneider}, {Schoorlemmer}, {Seglar Arroyo}, {Sinnis}, {Smith},
  {Springer}, {Surajbali}, {Tabachnick}, {Tanner}, {Tibolla}, {Tollefson},
  {Torres}, {Weisgarber}, {Westerhoff}, {Wood}, {Yapici}, {Zepeda}, {Zhou}, \&
  {HAWC Collaboration}}]{HAWC_crab_100TeV}
{Abeysekara}, A.~U., {Albert}, A., {Alfaro}, R., {et~al.} 2019, \apj, 881, 134,
  \dodoi{10.3847/1538-4357/ab2f7d}

\bibitem[{{Abramowski} {et~al.}(2014){Abramowski}, {Aharonian}, {Ait Benkhali},
  {Akhperjanian}, {Ang{\"u}ner}, {Backes}, {Balenderan}, {Balzer}, {Barnacka},
  {Becherini}, \& et~al.}]{HESS_diffuse_emission}
{Abramowski}, A., {Aharonian}, F., {Ait Benkhali}, F., {et~al.} 2014, Physical
  Review D: Particles, Fields, Gravitation and Cosmology, 90, 122007,
  \dodoi{10.1103/PhysRevD.90.122007}

\bibitem[{{Aharonian} {et~al.}(2006){Aharonian}, {Akhperjanian}, {Bazer-Bachi},
  {Beilicke}, {Benbow}, {Berge}, {Bernl{\"o}hr}, {Boisson}, {Bolz}, {Borrel},
  {Braun}, {Breitling}, {Brown}, {B{\"u}hler}, {B{\"u}sching}, {Carrigan},
  {Chadwick}, {Chounet}, {Cornils}, {Costamante}, {Degrange}, {Dickinson},
  {Djannati-Ata{\"i}}, {O'C.~Drury}, {Dubus}, {Egberts}, {Emmanoulopoulos},
  {Espigat}, {Feinstein}, {Ferrero}, {Fiasson}, {Fontaine}, {Funk}, {Funk},
  {Gallant}, {Giebels}, {Glicenstein}, {Goret}, {Hadjichristidis}, {Hauser},
  {Hauser}, {Heinzelmann}, {Henri}, {Hermann}, {Hinton}, {Hofmann}, {Holleran},
  {Horns}, {Jacholkowska}, {de Jager}, {Kh{\'e}lifi}, {Komin}, {Konopelko},
  {Kosack}, {Latham}, {Le Gallou}, {Lemi{\`e}re}, {Lemoine-Goumard}, {Lohse},
  {Martin}, {Martineau-Huynh}, {Marcowith}, {Masterson}, {McComb}, {de
  Naurois}, {Nedbal}, {Nolan}, {Noutsos}, {Orford}, {Osborne}, {Ouchrif},
  {Panter}, {Pelletier}, {Pita}, {P{\"u}hlhofer}, {Punch}, {Raubenheimer},
  {Raue}, {Rayner}, {Reimer}, {Reimer}, {Ripken}, {Rob}, {Rolland}, {Rowell},
  {Sahakian}, {Saug{\'e}}, {Schlenker}, {Schlickeiser}, {Schwanke}, {Sol},
  {Spangler}, {Spanier}, {Steenkamp}, {Stegmann}, {Superina}, {Tavernet},
  {Terrier}, {Th{\'e}oret}, {Tluczykont}, {van Eldik}, {Vasileiadis}, {Venter},
  {Vincent}, {V{\"o}lk}, {Wagner}, \& {Ward}}]{HESS_crab}
{Aharonian}, F., {Akhperjanian}, A.~G., {Bazer-Bachi}, A.~R., {et~al.} 2006,
  Astronomy and Astrophysics, 457, 899, \dodoi{10.1051/0004-6361:20065351}

\bibitem[{{Aharonian} {et~al.}(2009){Aharonian}, {Akhperjanian}, {Anton},
  {Barres de Almeida}, {Bazer-Bachi}, {Becherini}, {Behera}, {Benbow},
  {Bernl{\"o}hr}, {Boisson}, {Bochow}, {Borrel}, {Braun}, {Brion}, {Brucker},
  {Brun}, {B{\"u}hler}, {Bulik}, {B{\"u}sching}, {Boutelier}, {Carrigan},
  {Chadwick}, {Charbonnier}, {Chaves}, {Cheesebrough}, {Chounet}, {Clapson},
  {Coignet}, {Dalton}, {Daniel}, {Degrange}, {Deil}, {Dickinson},
  {Djannati-Ata{\"\i}}, {Domainko}, {O'C. Drury}, {Dubois}, {Dubus}, {Dyks},
  {Dyrda}, {Egberts}, {Emmanoulopoulos}, {Espigat}, {Farnier}, {Feinstein},
  {Fiasson}, {F{\"o}rster}, {Fontaine}, {F{\"u}{\ss}ling}, {Gabici}, {Gallant},
  {G{\'e}rard}, {Giebels}, {Glicenstein}, {Gl{\"u}ck}, {Goret}, {Hauser},
  {Hauser}, {Heinz}, {Heinzelmann}, {Henri}, {Hermann}, {Hinton}, {Hoffmann},
  {Hofmann}, {Holleran}, {Hoppe}, {Horns}, {Jacholkowska}, {de Jager}, {Jung},
  {Katarzy{\'n}ski}, {Katz}, {Kaufmann}, {Kendziorra}, {Kerschhaggl},
  {Khangulyan}, {Kh{\'e}lifi}, {Keogh}, {Komin}, {Kosack}, {Lamanna}, {Lenain},
  {Lohse}, {Marandon}, {Martin}, {Martineau-Huynh}, {Marcowith}, {Maurin},
  {McComb}, {Medina}, {Moderski}, {Moulin}, {Naumann-Godo}, {de Naurois},
  {Nedbal}, {Nekrassov}, {Niemiec}, {Nolan}, {Ohm}, {Olive}, {de O{\~n}a
  Wilhelmi}, {Orford}, {Ostrowski}, {Panter}, {Paz Arribas}, {Pedaletti},
  {Pelletier}, {Petrucci}, {Pita}, {P{\"u}hlhofer}, {Punch}, {Quirrenbach},
  {Raubenheimer}, {Raue}, {Rayner}, {Renaud}, {Reimer}, {Rieger}, {Ripken},
  {Rob}, {Rosier-Lees}, {Rowell}, {Rudak}, {Rulten}, {Ruppel}, {Sahakian},
  {Santangelo}, {Schlickeiser}, {Sch{\"o}ck}, {Schr{\"o}der}, {Schwanke},
  {Schwarzburg}, {Schwemmer}, {Shalchi}, {Skilton}, {Sol}, {Spangler},
  {Stawarz}, {Steenkamp}, {Stegmann}, {Superina}, {Tam}, {Tavernet}, {Terrier},
  {Tibolla}, {van Eldik}, {Vasileiadis}, {Venter}, {Venter}, {Vialle},
  {Vincent}, {Vivier}, {V{\"o}lk}, {Volpe}, {Wagner}, {Ward}, {Zdziarski}, \&
  {Zech}}]{HESSJ1908+063}
{Aharonian}, F., {Akhperjanian}, A.~G., {Anton}, G., {et~al.} 2009, \aap, 499,
  723, \dodoi{10.1051/0004-6361/200811357}

\bibitem[{{Ahnen} {et~al.}(2019){Ahnen}, {Ansoldi}, {Antonelli}, {Arcaro},
  {Baack}, {Babi{\'c}}, {}, {Banerjee}, {Bangale}, {Barres de Almeida},
  {Barrio}, {Becerra Gonz{\'a}lez}, {Bednarek}, {Bernardini}, {Berse}, {Berti},
  {Bhattacharyya}, {Biland}, {Blanch}, {Bonnoli}, {Carosi}, {Carosi},
  {Ceribella}, {Chatterjee}, {Colak}, {Colin}, {Colombo}, {Contreras},
  {Cortina}, {Covino}, {Cumani}, {da Vela}, {Dazzi}, {de Angelis}, {de Lotto},
  {Delfino}, {Delgado}, {di Pierro}, {Dom{\'\i}nguez}, {Dominis Prester},
  {Dorner}, {Doro}, {Einecke}, {Elsaesser}, {Fallah Ramazani},
  {Fern{\'a}ndez-Barral}, {Fidalgo}, {Fonseca}, {Font}, {Fruck}, {Galindo},
  {Garc{\'\i}a L{\'o}pez}, {Garczarczyk}, {Gaug}, {Giammaria}, {Godinovi{\'c}},
  {}, {Gora}, {Guberman}, {Hadasch}, {Hahn}, {Hassan}, {Hayashida}, {Herrera},
  {Hose}, {Hrupec}, {Ishio}, {Konno}, {Kubo}, {Kushida}, {Kuve{\v{z}}di{\'c}},
  {}, {Lelas}, {Lindfors}, {Lombardi}, {Longo}, {L{\'o}pez}, {Maggio},
  {Majumdar}, {Makariev}, {Maneva}, {Manganaro}, {Mannheim}, {Maraschi},
  {Mariotti}, {Mart{\'\i}nez}, {Masuda}, {Mazin}, {Mielke}, {Minev}, {Miranda},
  {Mirzoyan}, {Moralejo}, {Moreno}, {Moretti}, {Nagayoshi}, {Neustroev},
  {Niedzwiecki}, {Nievas Rosillo}, {Nigro}, {Nilsson}, {Ninci}, {Nishijima},
  {Noda}, {Nogu{\'e}s}, {Paiano}, {Palacio}, {Paneque}, {Paoletti}, {Paredes},
  {Pedaletti}, {Peresano}, {Persic}, {Prada Moroni}, {Prand ini}, {Puljak},
  {Garcia}, {Reichardt}, {Rhode}, {Rib{\'o}}, {Rico}, {Righi}, {Rugliancich},
  {Saito}, {Satalecka}, {Schweizer}, {Sitarek}, {{\v{S}}nidari{\'c}}, {},
  {Sobczynska}, {Stamerra}, {Strzys}, {Suri{\'c}}, {}, {Takahashi}, {Takalo},
  {Tavecchio}, {Temnikov}, {Terzi{\'c}}, {}, {Teshima}, {Torres-Alb{\`a}},
  {Treves}, {Tsujimoto}, {Vanzo}, {Vazquez Acosta}, {Vovk}, {Ward}, {Will},
  {Zari{\'c}}, {}, {MAGIC Collaboration}, {Albert}, {Alfaro}, {Alvarez},
  {Arceo}, {Arteaga-Vel{\'a}zquez}, {Avila Rojas}, {Ayala Solares}, {Becerril},
  {Belmont-Moreno}, {Benzvi}, {Bernal}, {Braun}, {Caballero-Mora},
  {Capistr{\'a}n}, {Carrami{\~n}ana}, {Casanova}, {Castillo}, {Cotti},
  {Cotzomi}, {Couti{\~n}o de Le{\'o}n}, {de Le{\'o}n}, {de La Fuente}, {Diaz
  Hernandez}, {Dichiara}, {Dingus}, {Duvernois}, {D{\'\i}az-V{\'e}lez},
  {Ellsworth}, {Engel}, {Enriquez-Rivera}, {Fiorino}, {Fleischhack}, {Fraija},
  {Garc{\'\i}a-Gonz{\'a}lez}, {Garfias}, {Gonz{\'a}lez-Mu{\~n}oz},
  {Gonz{\'a}lez}, {Goodman}, {Hampel-Arias}, {Harding}, {Hernand ez},
  {Hueyotl-Zahuantitla}, {Hui}, {H{\"u}ntemeyer}, {Iriarte}, {Jardin-Blicq},
  {Joshi}, {Kaufmann}, {Lara}, {Lauer}, {Lee}, {Lennarz}, {Le{\'o}n Vargas},
  {Linnemann}, {Longinotti}, {Luis-Raya}, {Luna-Garc{\'\i}a}, {L{\'o}pez-Coto},
  {Malone}, {Marinelli}, {Martinez}, {Martinez-Castellanos},
  {Mart{\'\i}nez-Castro}, {Mart{\'\i}nez-Huerta}, {Matthews},
  {Miranda-Romagnoli}, {Moreno}, {Mostaf{\'a}}, {Nayerhoda}, {Nellen},
  {Newbold}, {Nisa}, {Noriega-Papaqui}, {Pelayo}, {Pretz},
  {P{\'e}rez-P{\'e}rez}, {Ren}, {Rho}, {Rivi{\`e}re}, {Rosa-Gonz{\'a}lez},
  {Rosenberg}, {Ruiz-Velasco}, {Salesa Greus}, {Sandoval}, {Schneider}, {Seglar
  Arroyo}, {Sinnis}, {Smith}, {Springer}, {Surajbali}, {Taboada}, {Tibolla},
  {Tollefson}, {Torres}, {Ukwatta}, {Vianello}, {Villase{\~n}or}, {Werner},
  {Westerhoff}, {Wood}, {Yapici}, {Yodh}, {Zepeda}, {Zhou}, {{\'A}lvarez},
  {Hawc Collaboration}, {Ajello}, {Baldini}, {Barbiellini}, {Berenji},
  {Bissaldi}, {Bland ford}, {Bonino}, {Bottacini}, {Brandt}, {Bregeon},
  {Bruel}, {Cameron}, {Caputo}, {Caraveo}, {Castro}, {Cavazzuti}, {Chiaro},
  {Ciprini}, {Costantin}, {D'Ammando}, {de Palma}, {Desai}, {di Lalla}, {di
  Mauro}, {di Venere}, {Dom{\'\i}nguez}, {Favuzzi}, {Fukazawa}, {Funk},
  {Fusco}, {Gargano}, {Gasparrini}, {Giglietto}, {Giordano}, {Giroletti},
  {Glanzman}, {Green}, {Grenier}, {Guiriec}, {Harding}, {Hays}, {Hewitt},
  {Horan}, {J{\'o}hannesson}, {Kuss}, {Larsson}, {Liodakis}, {Longo},
  {Loparco}, {Lubrano}, {Magill}, {Maldera}, {Manfreda}, {Mazziotta}, {Mereu},
  {Michelson}, {Mizuno}, {Monzani}, {Morselli}, {Moskalenko}, {Negro}, {Nuss},
  {Omodei}, {Orienti}, {Orlando}, {Ormes}, {Palatiello}, {Paliya}, {Persic},
  {Pesce-Rollins}, {Petrosian}, {Piron}, {Porter}, {Principe}, {Rain{\`o}},
  {Rani}, {Razzano}, {Razzaque}, {Reimer}, {Reimer}, {Sgr{\`o}}, {Siskind},
  {Spandre}, {Spinelli}, {Tajima}, {Takahashi}, {Thayer}, {Thompson}, {Torres},
  {Torresi}, {Troja}, {Valverde}, {Wood}, {Yassine}, \& {Fermi-Lat
  Collaboration}}]{HAWC_Fermi_MAGIC}
{Ahnen}, M.~L., {Ansoldi}, S., {Antonelli}, L.~A., {et~al.} 2019, \mnras, 485,
  356, \dodoi{10.1093/mnras/stz089}

\bibitem[{{Albert} {et~al.}(2020){Albert}, {Alfaro}, {Alvarez}, {Camacho},
  {Arteaga-Vel{\'a}zquez}, {Arunbabu}, {Avila Rojas}, {Ayala Solares},
  {Baghmanyan}, {Belmont-Moreno}, {BenZvi}, {Brisbois}, {Caballero-Mora},
  {Capistr{\'a}n}, {Carrami{\~n}ana}, {Casanova}, {Cotti}, {Couti{\~n}o de
  Le{\'o}n}, {De la Fuente}, {Diaz Hernandez}, {Diaz-Cruz}, {Dingus},
  {DuVernois}, {Durocher}, {D{\'\i}az-V{\'e}lez}, {Ellsworth}, {Engel},
  {Espinoza}, {Fan}, {Fang}, {Alonso}, {Fleischhack}, {Fraija},
  {Galv{\'a}n-G{\'a}mez}, {Garcia}, {Garc{\'\i}a-Gonz{\'a}lez}, {Garfias},
  {Giacinti}, {Gonz{\'a}lez}, {Goodman}, {Harding}, {Hernandez}, {Hinton},
  {Hona}, {Huang}, {Hueyotl-Zahuantitla}, {H{\"u}ntemeyer}, {Iriarte},
  {Jardin-Blicq}, {Joshi}, {Kieda}, {Lara}, {Lee}, {Le{\'o}n Vargas},
  {Linnemann}, {Longinotti}, {Luis-Raya}, {Lundeen}, {L{\'o}pez-Coto},
  {Malone}, {Marandon}, {Martinez}, {Martinez-Castellanos},
  {Mart{\'\i}nez-Castro}, {Matthews}, {Miranda-Romagnoli}, {Morales-Soto},
  {Moreno}, {Mostaf{\'a}}, {Nayerhoda}, {Nellen}, {Newbold}, {Nisa},
  {Noriega-Papaqui}, {Olivera-Nieto}, {Omodei}, {Peisker}, {P{\'e}rez Araujo},
  {P{\'e}rez-P{\'e}rez}, {Ren}, {Rho}, {Rivi{\`e}re}, {Rosa-Gonz{\'a}lez},
  {Ruiz-Velasco}, {Salazar}, {Salesa Greus}, {Sandoval}, {Schneider},
  {Schoorlemmer}, {Serna}, {Sinnis}, {Smith}, {Springer}, {Surajbali},
  {Tollefson}, {Torres}, {Torres-Escobedo}, {Ukwatta}, {Ure{\~n}a-Mena},
  {Weisgarber}, {Werner}, {Willox}, {Zepeda}, {Zhou}, {de Le{\'o}n},
  {{\'A}lvarez}, \& {HAWC Collaboration}}]{3HWC_catalog}
{Albert}, A., {Alfaro}, R., {Alvarez}, C., {et~al.} 2020, \apj, 905, 76,
  \dodoi{10.3847/1538-4357/abc2d8}

\bibitem[{{Astropy Collaboration} {et~al.}(2013){Astropy Collaboration},
  {Robitaille}, {Tollerud}, {Greenfield}, {Droettboom}, {Bray}, {Aldcroft},
  {Davis}, {Ginsburg}, {Price-Whelan}, {Kerzendorf}, {Conley}, {Crighton},
  {Barbary}, {Muna}, {Ferguson}, {Grollier}, {Parikh}, {Nair}, {Unther},
  {Deil}, {Woillez}, {Conseil}, {Kramer}, {Turner}, {Singer}, {Fox}, {Weaver},
  {Zabalza}, {Edwards}, {Azalee Bostroem}, {Burke}, {Casey}, {Crawford},
  {Dencheva}, {Ely}, {Jenness}, {Labrie}, {Lim}, {Pierfederici}, {Pontzen},
  {Ptak}, {Refsdal}, {Servillat}, \& {Streicher}}]{astropy:2013}
{Astropy Collaboration}, {Robitaille}, T.~P., {Tollerud}, E.~J., {et~al.} 2013,
  \aap, 558, A33, \dodoi{10.1051/0004-6361/201322068}

\bibitem[{{Berge} {et~al.}(2007){Berge}, {Funk}, \& {Hinton}}]{HESS_background}
{Berge}, D., {Funk}, S., \& {Hinton}, J. 2007, Astronomy and Astrophysics, 466,
  1219, \dodoi{10.1051/0004-6361:20066674}

\bibitem[{{Cawley} {et~al.}(1985){Cawley}, {Fegan}, {Gibbs}, {Gorham},
  {Hillas}, {Lamb}, {Liebing}, {MacKeown}, {Porter}, {Stenger}, \&
  {Weekes}}]{Hillas_1985}
{Cawley}, M.~F., {Fegan}, D.~J., {Gibbs}, K., {et~al.} 1985, in International
  Cosmic Ray Conference, Vol.~3, 19th International Cosmic Ray Conference
  (ICRC19), Volume 3, 453--456

\bibitem[{{Cherenkov Telescope Array Consortium} {et~al.}(2019){Cherenkov
  Telescope Array Consortium}, {Acharya}, {Agudo}, {Al Samarai}, {Alfaro},
  {Alfaro}, {Alispach}, {Alves Batista}, {Amans}, \& {Amato}}]{CTAbook}
{Cherenkov Telescope Array Consortium}, {Acharya}, B.~S., {Agudo}, I., {et~al.}
  2019, {Science with the Cherenkov Telescope Array}, \dodoi{10.1142/10986}

\bibitem[{{Cordes} {et~al.}(2006){Cordes}, {Freire}, {Lorimer}, {Camilo},
  {Champion}, {Nice}, {Ramachandran}, {Hessels}, {Vlemmings}, {van Leeuwen},
  {Ransom}, {Bhat}, {Arzoumanian}, {McLaughlin}, {Kaspi}, {Kasian}, {Deneva},
  {Reid}, {Chatterjee}, {Han}, {Backer}, {Stairs}, {Deshpande}, \&
  {Faucher-Gigu{\`e}re}}]{Discovery_PSRJ1928}
{Cordes}, J.~M., {Freire}, P.~C.~C., {Lorimer}, D.~R., {et~al.} 2006, The
  Astronomical Journal, 637, 446, \dodoi{10.1086/498335}

\bibitem[{{Dame} {et~al.}(2001){Dame}, {Hartmann}, \& {Thaddeus}}]{COsurvey}
{Dame}, T.~M., {Hartmann}, D., \& {Thaddeus}, P. 2001, The Astrophysical
  Journal, 547, 792, \dodoi{10.1086/318388}

\bibitem[{{de Naurois} \& {Rolland}(2009)}]{Modelpp}
{de Naurois}, M., \& {Rolland}, L. 2009, Astroparticle Physics, 32, 231,
  \dodoi{10.1016/j.astropartphys.2009.09.001}

\bibitem[{{Deil} {et~al.}(2017){Deil}, {Zanin}, {Lefaucheur}, {Boisson},
  {Khelifi}, {Terrier}, {Wood}, {Mohrmann}, {Chakraborty}, {Watson},
  {Lopez-Coto}, {Klepser}, {Cerruti}, {Lenain}, {Acero}, {Djannati-Ata{\"\i}},
  {Pita}, {Bosnjak}, {Trichard}, {Vuillaume}, {Donath}, {Consortium}, {King},
  {Jouvin}, {Owen}, {Sipocz}, {Lennarz}, {Voruganti}, {Spir-Jacob}, {Ruiz}, \&
  {Arribas}}]{gammapy:2017}
{Deil}, C., {Zanin}, R., {Lefaucheur}, J., {et~al.} 2017, in International
  Cosmic Ray Conference, Vol. 301, 35th International Cosmic Ray Conference
  (ICRC2017), 766.
\newblock \doarXiv{1709.01751}

\bibitem[{{Fuerst} {et~al.}(1987){Fuerst}, {Reich}, {Reich}, {Handa}, \&
  {Sofue}}]{SNRG42.8+0.6}
{Fuerst}, E., {Reich}, W., {Reich}, P., {Handa}, T., \& {Sofue}, Y. 1987,
  \aaps, 69, 403

\bibitem[{{H.E.S.S. Collaboration} {et~al.}(2018{\natexlab{a}}){H.E.S.S.
  Collaboration}, {Abdalla}, {Abramowski}, {Aharonian}, {Ait Benkhali},
  {Ang{\"u}ner}, {Arakawa}, {Arrieta}, {Aubert}, {Backes}, {Balzer}, {Barnard},
  {Becherini}, {Becker Tjus}, {Berge}, {Bernhard}, {Bernl{\"o}hr}, {Blackwell},
  {B{\"o}ttcher}, {Boisson}, {Bolmont}, {Bonnefoy}, {Bordas}, {Bregeon},
  {Brun}, {Brun}, {Bryan}, {B{\"u}chele}, {Bulik}, {Capasso}, {Carrigan},
  {Caroff}, {Carosi}, {Casanova}, {Cerruti}, {Chakraborty}, {Chaves}, {Chen},
  {Chevalier}, {Colafrancesco}, {Condon}, {Conrad}, {Davids}, {Decock}, {Deil},
  {Devin}, {deWilt}, {Dirson}, {Djannati-Ata{\"\i}}, {Domainko}, {Donath},
  {Drury}, {Dutson}, {Dyks}, {Edwards}, {Egberts}, {Eger}, {Emery},
  {Ernenwein}, {Eschbach}, {Farnier}, {Fegan}, {Fernandes}, {Fiasson},
  {Fontaine}, {F{\"o}rster}, {Funk}, {F{\"u}{\ss}ling}, {Gabici}, {Gallant},
  {Garrigoux}, {Gast}, {Gat{\'e}}, {Giavitto}, {Giebels}, {Glawion},
  {Glicenstein}, {Gottschall}, {Grondin}, {Hahn}, {Haupt}, {Hawkes},
  {Heinzelmann}, {Henri}, {Hermann}, {Hinton}, {Hofmann}, {Hoischen}, {Holch},
  {Holler}, {Horns}, {Ivascenko}, {Iwasaki}, {Jacholkowska}, {Jamrozy},
  {Jankowsky}, {Jankowsky}, {Jingo}, {Jouvin}, {Jung-Richardt}, {Kastendieck},
  {Katarzy{\'n}ski}, {Katsuragawa}, {Katz}, {Kerszberg}, {Khangulyan},
  {Kh{\'e}lifi}, {King}, {Klepser}, {Klochkov}, {Klu{\'z}niak}, {Komin},
  {Kosack}, {Krakau}, {Kraus}, {Kr{\"u}ger}, {Laffon}, {Lamanna}, {Lau},
  {Lees}, {Lefaucheur}, {Lemi{\`e}re}, {Lemoine-Goumard}, {Lenain}, {Leser},
  {Lohse}, {Lorentz}, {Liu}, {L{\'o}pez-Coto}, {Lypova}, {Marandon},
  {Malyshev}, {Marcowith}, {Mariaud}, {Marx}, {Maurin}, {Maxted}, {Mayer},
  {Meintjes}, {Meyer}, {Mitchell}, {Moderski}, {Mohamed}, {Mohrmann},
  {Mor{\r{a}}}, {Moulin}, {Murach}, {Nakashima}, {de Naurois}, {Ndiyavala},
  {Niederwanger}, {Niemiec}, {Oakes}, {O'Brien}, {Odaka}, {Ohm}, {Ostrowski},
  {Oya}, {Padovani}, {Panter}, {Parsons}, {Paz Arribas}, {Pekeur}, {Pelletier},
  {Perennes}, {Petrucci}, {Peyaud}, {Piel}, {Pita}, {Poireau}, {Poon},
  {Prokhorov}, {Prokoph}, {P{\"u}hlhofer}, {Punch}, {Quirrenbach}, {Raab},
  {Rauth}, {Reimer}, {Reimer}, {Renaud}, {de los Reyes}, {Rieger}, {Rinchiuso},
  {Romoli}, {Rowell}, {Rudak}, {Rulten}, {Safi-Harb}, {Sahakian}, {Saito},
  {Sanchez}, {Santangelo}, {Sasaki}, {Schandri}, {Schlickeiser},
  {Sch{\"u}ssler}, {Schulz}, {Schwanke}, {Schwemmer}, {Seglar-Arroyo},
  {Settimo}, {Seyffert}, {Shafi}, {Shilon}, {Shiningayamwe}, {Simoni}, {Sol},
  {Spanier}, {Spir-Jacob}, {Stawarz}, {Steenkamp}, {Stegmann}, {Steppa},
  {Sushch}, {Takahashi}, {Tavernet}, {Tavernier}, {Taylor}, {Terrier},
  {Tibaldo}, {Tiziani}, {Tluczykont}, {Trichard}, {Tsirou}, {Tsuji}, {Tuffs},
  {Uchiyama}, {van der Walt}, {van Eldik}, {van Rensburg}, {van Soelen},
  {Vasileiadis}, {Veh}, {Venter}, {Viana}, {Vincent}, {Vink}, {Voisin},
  {V{\"o}lk}, {Vuillaume}, {Wadiasingh}, {Wagner}, {Wagner}, {Wagner}, {White},
  {Wierzcholska}, {Willmann}, {W{\"o}rnlein}, {Wouters}, {Yang}, {Zaborov},
  {Zacharias}, {Zanin}, {Zdziarski}, {Zech}, {Zefi}, {Ziegler}, {Zorn}, \&
  {{\.Z}ywucka}}]{HGPS}
{H.E.S.S. Collaboration}, {Abdalla}, H., {Abramowski}, A., {et~al.}
  2018{\natexlab{a}}, \aap, 612, A1, \dodoi{10.1051/0004-6361/201732098}

\bibitem[{{H.E.S.S. Collaboration} {et~al.}(2018{\natexlab{b}}){H.E.S.S.
  Collaboration}, {Abdalla}, {Abramowski}, {Aharonian}, {Ait Benkhali},
  {Ang{\"u}ner}, {Arakawa}, {Arrieta}, {Aubert}, {Backes}, {Balzer}, {Barnard},
  {Becherini}, {Becker Tjus}, {Berge}, {Bernhard}, {Bernl{\"o}hr}, {Blackwell},
  {B{\"o}ttcher}, {Boisson}, {Bolmont}, {Bonnefoy}, {Bordas}, {Bregeon},
  {Brun}, {Brun}, {Bryan}, {B{\"u}chele}, {Bulik}, {Capasso}, {Caroff},
  {Carosi}, {Casanova}, {Cerruti}, {Chakraborty}, {Chaves}, {Chen},
  {Chevalier}, {Colafrancesco}, {Condon}, {Conrad}, {Davids}, {Decock}, {Deil},
  {Devin}, {deWilt}, {Dirson}, {Djannati-Ata{\"\i}}, {Donath}, {Drury},
  {Dutson}, {Dyks}, {Edwards}, {Egberts}, {Emery}, {Ernenwein}, {Eschbach},
  {Farnier}, {Fegan}, {Fernandes}, {Fernand ez}, {Fiasson}, {Fontaine}, {Funk},
  {F{\"u}{\ss}ling}, {Gabici}, {Gallant}, {Garrigoux}, {Gat{\'e}}, {Giavitto},
  {Giebels}, {Glawion}, {Glicenstein}, {Gottschall}, {Grondin}, {Hahn},
  {Haupt}, {Hawkes}, {Heinzelmann}, {Henri}, {Hermann}, {Hinton}, {Hofmann},
  {Hoischen}, {Holch}, {Holler}, {Horns}, {Ivascenko}, {Iwasaki},
  {Jacholkowska}, {Jamrozy}, {Jankowsky}, {Jankowsky}, {Jingo}, {Jouvin},
  {Jung-Richardt}, {Kastendieck}, {Katarzy{\'n}ski}, {Katsuragawa}, {Katz},
  {Kerszberg}, {Khangulyan}, {Kh{\'e}lifi}, {King}, {Klepser}, {Klochkov},
  {Klu{\'z}niak}, {Komin}, {Kosack}, {Krakau}, {Kraus}, {Kr{\"u}ger}, {Laffon},
  {Lamanna}, {Lau}, {Lees}, {Lefaucheur}, {Lemi{\`e}re}, {Lemoine-Goumard},
  {Lenain}, {Leser}, {Lohse}, {Lorentz}, {Liu}, {L{\'o}pez-Coto}, {Lypova},
  {Malyshev}, {Marandon}, {Marcowith}, {Mariaud}, {Marx}, {Maurin}, {Maxted},
  {Mayer}, {Meintjes}, {Meyer}, {Mitchell}, {Moderski}, {Mohamed}, {Mohrmann},
  {Mor{\r{a}}}, {Moulin}, {Murach}, {Nakashima}, {de Naurois}, {Ndiyavala},
  {Niederwanger}, {Niemiec}, {Oakes}, {O'Brien}, {Odaka}, {Ohm}, {Ostrowski},
  {Oya}, {Padovani}, {Panter}, {Parsons}, {Pekeur}, {Pelletier}, {Perennes},
  {Petrucci}, {Peyaud}, {Piel}, {Pita}, {Poireau}, {Poon}, {Prokhorov},
  {Prokoph}, {P{\"u}hlhofer}, {Punch}, {Quirrenbach}, {Raab}, {Rauth},
  {Reimer}, {Reimer}, {Renaud}, {de los Reyes}, {Rieger}, {Rinchiuso},
  {Romoli}, {Rowell}, {Rudak}, {Rulten}, {Safi-Harb}, {Sahakian}, {Saito},
  {Sanchez}, {Santangelo}, {Sasaki}, {Schlickeiser}, {Sch{\"u}ssler}, {Schulz},
  {Schwanke}, {Schwemmer}, {Seglar-Arroyo}, {Settimo}, {Seyffert}, {Shafi},
  {Shilon}, {Shiningayamwe}, {Simoni}, {Sol}, {Spanier}, {Spir-Jacob},
  {Stawarz}, {Steenkamp}, {Stegmann}, {Steppa}, {Sushch}, {Takahashi},
  {Tavernet}, {Tavernier}, {Taylor}, {Terrier}, {Tibaldo}, {Tiziani},
  {Tluczykont}, {Trichard}, {Tsirou}, {Tsuji}, {Tuffs}, {Uchiyama}, {van der
  Walt}, {van Eldik}, {van Rensburg}, {van Soelen}, {Vasileiadis}, {Veh},
  {Venter}, {Viana}, {Vincent}, {Vink}, {Voisin}, {V{\"o}lk}, {Vuillaume},
  {Wadiasingh}, {Wagner}, {Wagner}, {Wagner}, {White}, {Wierzcholska},
  {Willmann}, {W{\"o}rnlein}, {Wouters}, {Yang}, {Zaborov}, {Zacharias},
  {Zanin}, {Zdziarski}, {Zech}, {Zefi}, {Ziegler}, {Zorn}, \&
  {{\.Z}ywucka}}]{HESS_W44}
---. 2018{\natexlab{b}}, \aap, 612, A3, \dodoi{10.1051/0004-6361/201732125}

\bibitem[{{HI4PI Collaboration} {et~al.}(2016){HI4PI Collaboration}, {Ben
  Bekhti, N.}, {Fl\"oer, L.}, {Keller, R.}, {Kerp, J.}, {Lenz, D.}, {Winkel,
  B.}, {Bailin, J.}, {Calabretta, M. R.}, {Dedes, L.}, {Ford, H. A.}, {Gibson,
  B. K.}, {Haud, U.}, {Janowiecki, S.}, {Kalberla, P. M. W.}, {Lockman, F. J.},
  {McClure-Griffiths, N. M.}, {Murphy, T.}, {Nakanishi, H.}, {Pisano, D. J.},
  \& {Staveley-Smith, L.}}]{HI4PI}
{HI4PI Collaboration}, {Ben Bekhti, N.}, {Fl\"oer, L.}, {et~al.} 2016, A\&A,
  594, A116, \dodoi{10.1051/0004-6361/201629178}

\bibitem[{{Hnatyk} {et~al.}(2020){Hnatyk}, {Hnatyk}, {Zhdanov}, \&
  {Voitsekhovskyi}}]{SGR1900+14}
{Hnatyk}, B., {Hnatyk}, R., {Zhdanov}, V., \& {Voitsekhovskyi}, V. 2020, arXiv
  e-prints, arXiv:2009.06081.
\newblock \doarXiv{2009.06081}

\bibitem[{Jardin-Blicq(2019)}]{MyThesis}
Jardin-Blicq, A. 2019, PhD thesis, Heidelberg.
\newblock \url{http://hdl.handle.net/21.11116/0000-0004-F359-6}

\bibitem[{{Kargaltsev} {et~al.}(2012){Kargaltsev}, {Durant}, {Pavlov}, \&
  {Garmire}}]{Chandra_10ks_J1928}
{Kargaltsev}, O., {Durant}, M., {Pavlov}, G.~G., \& {Garmire}, G. 2012, \apjs,
  201, 37, \dodoi{10.1088/0067-0049/201/2/37}

\bibitem[{{Lazarus} {et~al.}(2015){Lazarus}, {Brazier}, {Hessels},
  {Karako-Argaman}, {Kaspi}, {Lynch}, {Madsen}, {Patel}, {Ransom}, {Scholz},
  {Swiggum}, {Zhu}, {Allen}, {Bogdanov}, {Camilo}, {Cardoso}, {Chatterjee},
  {Cordes}, {Crawford}, {Deneva}, {Ferdman}, {Freire}, {Jenet}, {Knispel},
  {Lee}, {van Leeuwen}, {Lorimer}, {Lyne}, {McLaughlin}, {Siemens}, {Spitler},
  {Stairs}, {Stovall}, \& {Venkataraman}}]{Arecibo_pulsar_survey}
{Lazarus}, P., {Brazier}, A., {Hessels}, J.~W.~T., {et~al.} 2015, \apj, 812,
  81, \dodoi{10.1088/0004-637X/812/1/81}

\bibitem[{{Lopez-Coto} {et~al.}(2017){Lopez-Coto}, {Marandon}, \&
  {Brun}}]{ICRC_J1928}
{Lopez-Coto}, R., {Marandon}, V., \& {Brun}, F. 2017, in International Cosmic
  Ray Conference, Vol. 301, 35th International Cosmic Ray Conference
  (ICRC2017), 732.
\newblock \doarXiv{1708.03137}

\bibitem[{{Mazets} {et~al.}(1979){Mazets}, {Golenetskij}, \&
  {Guryan}}]{SGR1900}
{Mazets}, E.~P., {Golenetskij}, S.~V., \& {Guryan}, Y.~A. 1979, Soviet
  Astronomy Letters, 5, 343

\bibitem[{Mohrmann {et~al.}(2019)Mohrmann, Specovius, Tiziani, Funk, Malyshev,
  Nakashima, \& van Eldik}]{Lars_bkg_model}
Mohrmann, L., Specovius, A., Tiziani, D., {et~al.} 2019, Astronomy \&
  Astrophysics, 632, A72, \dodoi{10.1051/0004-6361/201936452}

\bibitem[{{Mori} {et~al.}(2020){Mori}, {An}, {Feng}, {Malone}, {Prado},
  {Schutt}, {Dingus}, {Gotthelf}, {Hailey}, {Hare}, {Kargaltsev}, \&
  {Mukherjee}}]{J1928_dark_accelerator}
{Mori}, K., {An}, H., {Feng}, Q., {et~al.} 2020, \apj, 897, 129,
  \dodoi{10.3847/1538-4357/ab9631}

\bibitem[{{Nigro} {et~al.}(2019){Nigro}, {Deil}, {Zanin}, {Hassan}, {King},
  {Ruiz}, {Saha}, {Terrier}, {Br{\"u}gge}, {N{\"o}the}, {Bird}, {Lin},
  {Aleksi{\'c}}, {Boisson}, {Contreras}, {Donath}, {Jouvin}, {Kelley-Hoskins},
  {Khelifi}, {Kosack}, {Rico}, \& {Sinha}}]{gammapy:2019}
{Nigro}, C., {Deil}, C., {Zanin}, R., {et~al.} 2019, \aap, 625, A10,
  \dodoi{10.1051/0004-6361/201834938}

\bibitem[{{Parsons} \& {Hinton}(2014)}]{ImPACT}
{Parsons}, R.~D., \& {Hinton}, J.~A. 2014, Astroparticle Physics, 56, 26,
  \dodoi{10.1016/j.astropartphys.2014.03.002}

\bibitem[{{Price-Whelan} {et~al.}(2018){Price-Whelan}, {Sip{\H{o}}cz},
  {G{\"u}nther}, {Lim}, {Crawford}, {Conseil}, {Shupe}, {Craig}, {Dencheva},
  {Ginsburg}, {VanderPlas}, {Bradley}, {P{\'e}rez-Su{\'a}rez}, {de Val-Borro},
  {Paper Contributors}, {Aldcroft}, {Cruz}, {Robitaille}, {Tollerud},
  {Coordination Committee}, {Ardelean}, {Babej}, {Bach}, {Bachetti}, {Bakanov},
  {Bamford}, {Barentsen}, {Barmby}, {Baumbach}, {Berry}, {Biscani}, {Boquien},
  {Bostroem}, {Bouma}, {Brammer}, {Bray}, {Breytenbach}, {Buddelmeijer},
  {Burke}, {Calderone}, {Cano Rodr{\'\i}guez}, {Cara}, {Cardoso}, {Cheedella},
  {Copin}, {Corrales}, {Crichton}, {D{\textquoteright}Avella}, {Deil},
  {Depagne}, {Dietrich}, {Donath}, {Droettboom}, {Earl}, {Erben}, {Fabbro},
  {Ferreira}, {Finethy}, {Fox}, {Garrison}, {Gibbons}, {Goldstein}, {Gommers},
  {Greco}, {Greenfield}, {Groener}, {Grollier}, {Hagen}, {Hirst}, {Homeier},
  {Horton}, {Hosseinzadeh}, {Hu}, {Hunkeler}, {Ivezi{\'c}}, {Jain}, {Jenness},
  {Kanarek}, {Kendrew}, {Kern}, {Kerzendorf}, {Khvalko}, {King}, {Kirkby},
  {Kulkarni}, {Kumar}, {Lee}, {Lenz}, {Littlefair}, {Ma}, {Macleod},
  {Mastropietro}, {McCully}, {Montagnac}, {Morris}, {Mueller}, {Mumford},
  {Muna}, {Murphy}, {Nelson}, {Nguyen}, {Ninan}, {N{\"o}the}, {Ogaz}, {Oh},
  {Parejko}, {Parley}, {Pascual}, {Patil}, {Patil}, {Plunkett}, {Prochaska},
  {Rastogi}, {Reddy Janga}, {Sabater}, {Sakurikar}, {Seifert}, {Sherbert},
  {Sherwood-Taylor}, {Shih}, {Sick}, {Silbiger}, {Singanamalla}, {Singer},
  {Sladen}, {Sooley}, {Sornarajah}, {Streicher}, {Teuben}, {Thomas},
  {Tremblay}, {Turner}, {Terr{\'o}n}, {van Kerkwijk}, {de la Vega}, {Watkins},
  {Weaver}, {Whitmore}, {Woillez}, {Zabalza}, \& {Contributors}}]{astropy:2018}
{Price-Whelan}, A.~M., {Sip{\H{o}}cz}, B.~M., {G{\"u}nther}, H.~M., {et~al.}
  2018, \aj, 156, 123, \dodoi{10.3847/1538-3881/aabc4f}

\bibitem[{{Schoorlemmer}(2019)}]{SWGO_ICRC2019}
{Schoorlemmer}, H. 2019, in International Cosmic Ray Conference, Vol.~36, 36th
  International Cosmic Ray Conference (ICRC2019), 785.
\newblock \doarXiv{1908.08858}

\bibitem[{{Vall{\'e}e}(2014)}]{spiral_arms}
{Vall{\'e}e}, J.~P. 2014, The Astronomical Journal Suplement, 215, 1,
  \dodoi{10.1088/0067-0049/215/1/1}

\bibitem[{{Vianello} {et~al.}(2015){Vianello}, {Lauer}, {Younk}, {Tibaldo},
  {Burgess}, {Ayala}, {Harding}, {Hui}, {Omodei}, \& {Zhou}}]{3ML}
{Vianello}, G., {Lauer}, R.~J., {Younk}, P., {et~al.} 2015, Proceedings of the
  34th ICRC, The Hague, Netherlands.
\newblock \doarXiv{1507.08343}

\bibitem[{{Wakely} \& {Horan}(2008)}]{TeVCat}
{Wakely}, S.~P., \& {Horan}, D. 2008, in International Cosmic Ray Conference,
  Vol.~3, International Cosmic Ray Conference, 1341--1344

\bibitem[{Wilks(1938)}]{Wilks}
Wilks, S.~S. 1938, Ann. Math. Statist., 9, 60, \dodoi{10.1214/aoms/1177732360}

\bibitem[{{Yao} {et~al.}(2017){Yao}, {Manchester}, \& {Wang}}]{ymw17}
{Yao}, J.~M., {Manchester}, R.~N., \& {Wang}, N. 2017, \apj, 835, 29,
  \dodoi{10.3847/1538-4357/835/1/29}

\end{thebibliography}
\bibliographystyle{aasjournal}

%% This command is needed to show the entire author+affiliation list when
%% the collaboration and author truncation commands are used.  It has to
%% go at the end of the manuscript.
%\allauthors

%% Include this line if you are using the \added, \replaced, \deleted
%% commands to see a summary list of all changes at the end of the article.
%\listofchanges

\end{document}